%% file: main.tex
\documentclass{article}
\usepackage[utf8]{inputenc}
\pdfoutput=1

\usepackage{hyperref,xcolor}

\usepackage{amsmath, amsfonts, amssymb}
\usepackage{cite}
\usepackage[super,sort&compress]{natbib}       
\usepackage{arydshln}
\usepackage{float}
\usepackage{authblk}
\usepackage{geometry}

\usepackage{listings} 
 
\usepackage{rotating,graphicx,subcaption} 

\def\E{\mbox{\rm E}}

\def\R{\mbox{\rm R}}

\usepackage[colorinlistoftodos, textwidth=42mm, shadow]{todonotes}

\setcitestyle{citesep={,}}

\begin{document}

\title{A comparative study to alternatives to the log-rank test}

\author[1]{Ina Dormuth}
\author[2]{Tiantian Liu}
\author[3]{Jin Xu}
\author[1,4]{Markus Pauly}
\author[5]{Marc Ditzhaus}

\newcommand{\mbf}{\bm}
\newcommand{\trans}{'}
\def\R{{ \mathbb{R} }}
\def\N{{ \mathbb{N} }}
\def\E{ E }
\def\I{\mathbf{1}}
\def\P{ P }


\affil[1]{Department of Statistics, TU Dortmund University, Dortmund, Germany}
\affil[2]{Technion – Israel Institute of Technology, Haifa, Israel}    
\affil[3]{East China Normal University, Shanghai, China}    
\affil[4]{Research Center Trustworthy Data Science and Security, UA Ruhr, Dortmund}
\affil[5]{Department of Mathematics,  Otto von Guericke University Magdeburg, Magdeburg, Germany}
\maketitle

\begin{abstract}
	Studies to compare the survival of two or more groups using time-to-event data are of high importance in medical research. The gold standard  
	is the log-rank test, which is optimal under proportional hazards. 
	As the latter is no simple regularity assumption, we are interested in evaluating the power of various statistical tests under different settings including proportional and non-proportional hazards with a special emphasize on crossing hazards. This challenge has been going on for many years now and multiple methods have already been investigated in extensive simulation studies. However, in recent years new omnibus tests and methods based on the restricted mean survival time appeared that have been strongly recommended in biometric literature. Thus, to give updated recommendations, we perform a vast simulation study 
	to compare tests that showed high power in previous studies with these more recent approaches. We thereby analyze various simulation settings with varying survival and censoring distributions, unequal censoring between groups, small sample sizes and unbalanced group sizes. Overall, omnibus tests are more robust in terms of power against deviations from the proportional hazards assumption.
	
\end{abstract}

\vskip .5in

\noindent {\bf KEYWORDS:} survival analysis, crossing hazards, non-proportional hazards, simulation study, log-rank


\input{1_Introduction}
\input{2_Methods}

\input{3_Simulations}

\input{4_Data_Example}

\input{5_Discussion}

\input{6_Conclusion}

\bibliography{biblio}
\bibliographystyle{ama}

\newpage
\input{7_Appendix}

\end{document}

%% file: 1_Introduction.tex
\section*{Introduction}

The distributional comparison of two populations with censored time-to-event data is one of the most common inferential problems in survival analysis. The log-rank test is used as a standard tool in many medical or clinical studies. It is known to be optimal under the assumption of proportional hazards (PH). 
However, this assumption is often not met in reality due to various forms of derivation such as
crossing hazards, or early/late differences in survival curves. Kristiansen \cite{kristiansen2012} conducted a survey revealing that in $70\%$ of studies with crossing survival curves the log-rank test was used even though this leads to loss in power. Furthermore, Trinquart et al. \cite{trinquart2016comparison} revisited $54$ phase III oncology studies from five leading medical journals (New England Journal of Medicine, Lancet, Lancet Oncology, Journal of Clinical Oncology, Journal of the
American Medical Association) and found that for almost a fourth of the comparisons the proportional hazard assumption was rejected.  Non-proportionality as severe as crossing can appear when the treatment effects change over time. A common example is seen in immunotherapy which bears an early high risk but a long-term benefit. \cite{mick2015statistical, alexander2018hazards} Thus, the question on how to deal with non-proportional hazards is of high interest and has been investigated by many authors. 
For example, Royston and Parmar \cite{roystonSimulationStudyComparing2020} published a simulation study comparing nine methods implemented in Stata and showed a preference for modified weighted log-rank tests \cite{fleming1987supremum, royston2014approach, lee1996some, karrison2016versatile}. However, they did not include the situation of crossing hazards. Another overview was given by Lin et al. \cite{lin2020alternative}, focusing on combined weighted Kaplan-Meier and weighted log-rank tests. They conclude that as long as we do not have prior knowledge the MaxCombo test showed the most robust behavior among the tests under consideration. \cite{lin2020alternative}
Perhaps the most extensive study regarding crossing hazards was given in Li et al. \cite{liStatisticalInferenceMethods2015} who compared $21$ tests designed to handle crossing hazards. They stated that the two-stage test by Qiu and Sheng \cite{qiu2008} or the test by Kraus \cite{Kraus} are the most suitable among the studied tests. A general overview of existing methods and recommendations regarding trial design was created by Ananthakrishnan et al. \cite{ananthakrishnanCriticalReviewOncology2021} without numerical comparison.
None of the mentioned reviews considered new results on 
projection type, sample space partition or area under the survival curve tests. \cite{BrendelETAL2014,DitzhausFriedrich2018,gorfineKsampleOmnibusNonproportional2019,liuResamplingBasedTest2020} Recently, some of the new procedures have shown considerable power advantages in illustrative data analyses. \cite{dormuth2022}

We therefore enrich these investigations  by comparing the best performers from the above already existing simulation studies with more recent approaches. Our comprehensive simulation study covers $20$ representative scenarios including four null scenarios, four scenarios with PH, four scenarios with non-PH (excluding crossing structures) and eight scenarios with a special emphasize on crossing hazards. Since most procedures exhibit good
properties for large samples, our study focuses on small to moderate sample sizes.
In the next section we will review more details on the tests under study and their implementation. Afterwards, the different simulation and parameter settings are presented alongside with the results of the simulation study. The utility of the tests is further evaluated using reconstructed data from a phase~III clinical trial with moderate sample size. The findings are then discussed and conclusions are drawn, particularly focusing on the tests' power. 

%% file: 2_Methods.tex
\section*{Methods}
Multiple approaches to test the hypothesis of two equal survival functions have been developed. For ease of presentation, we categorize them in four groups and review the main ideas of the recommended ones in each group. Details on the methods can be found in the cited literature as well as the extended methods section in the Supplement.


\subsubsection*{Log-rank test and its weighted variants}
The standard to compare two survival functions $S_1$ and $S_2$ is the log-rank test (LR).\cite{singh2011survival} It belongs to the class of weighted log-rank tests \cite{fleming2011counting} that use the difference between the expected and observed number of events to derive a test statistic. These tests differ in the weight functions that they are employing. For instance, the log-rank test gives the same weight to all event times. Therefore, it is optimal under proportional hazards. The Peto-Peto test (PP) uses the Kaplan-Meier estimator $\widehat{S}(t)$ of the survival function as weight, which leads to a test that is more sensitive to early differences.\cite{legrand2021advanced} Various approaches to compute sample sizes for log-rank tests have been introduced, with Schoenfeld’s formula being the most popular.\cite{schoenfeld1983sample} 

In reality, due to the lack of prior information about the survival behavior of comparing populations, any mismatch of weight (or test) selection and difference in the true survival functions will lead to sub-optimal power performance. \cite{liStatisticalInferenceMethods2015}

\subsubsection*{Two-stage test}

The two-stage (TS) method introduced by Qiu and Sheng \cite{qiu2008} provides a solution to the weight selection problem in dealing with possible non-PH situations. The procedure gets its name from the sequential testing approach. More specifically, it conducts the standard log-rank test in the first stage. If the LR test does not reject the null hypothesis, an asymptotically independent test for crossing hazards is carried out. It is shown to be efficient with good adaptation and reliable in power performance under both PH and non-PH situations. \cite{qiu2008,liuResamplingBasedTest2020} The approach was extended to the k-sample case employing asymptotically independent tests. \cite{chenComparisonMultipleHazard2016}

\subsubsection*{Omnibus tests}

Another remedy to avoid potentially sub-optimal power performance 
is to use an omnibus test that does not have any inclination of the alternative hypothesis.

The mdir test proposed by Brendel et al.\cite{BrendelETAL2014} and revisited by Ditzhaus and Friedrich \cite{DitzhausFriedrich2018} uses a quadratic form-type statistic in multiple weighted LR statistics to cover broader alternatives. The test has high power for all alternatives corresponding to the chosen weights and combinations thereof. The test should be used especially when no prior information is available, because with prior knowledge a weighted test with only one suitable weight would have a higher power. A notable feature of the mdir test is that its permuted version allows to handle small sample cases with satisfactory type-I error and power performance.\cite{BrendelETAL2014,DitzhausFriedrich2018} The mdir test was extended to handle the one-sided testing problem as well as factorial designs \cite{ditzhaus2019wild, ditzhaus2021casanova}. 
A procedure for sample size calculation does not exist yet.

The class of maximum weighted log-rank tests bears a different approach to combine multiple weighted log-rank tests. Here, multiple test statistics with different weights are considered and the final test statistic is defined as the maximum over all of them. The MaxCombo test (MC) proposed by Lin et al. \cite{lin2020alternative} combines four weighted log-rank tests with Flemming-Harrington type weights targeting difference in survival functions with PH, late difference, middle difference, and early difference, respectively. An iterative sample size calculation approach was provided by Roychoudhury et al. \cite{roychoudhury2021robust}. The test can also be used for one-sided hypotheses. 

Gorfine et al.\cite{gorfineKsampleOmnibusNonproportional2019} introduced K-sample omnibus non-proportional
hazards (KONP) tests based on sample space partition that also tackles right censored data. P-values are obtained employing a censoring-friendly permutation procedure. The provided tests are based on two different test statistics, namely the log-likelihood ratio (KONP\_llr) and the chi-squared test statistic (KONP\_chi). Extensive simulation studies\cite{gorfineKsampleOmnibusNonproportional2019} showed that the choice of test statistic does not influence the performance. Hence, we only consider the KONP\_chi test in our study. 

\subsubsection*{Tests based on the area under the survival curve}
Tests based on restricted mean survival times (RMST) are often advocated in the context of crossing hazards. \cite{royston2011use, roystonRestrictedMeanSurvival2013, kim2017restricted, trinquart2016comparison} The RMST can be interpreted as the mean of event-free survival time up to $\tau$, where $\tau$ is a pre-defined time till which the truncated mean is of interest. In practice, $\tau$ is recommended to be $90\%$ of the minimum of the largest censored or uncensored event-time in the two groups. \cite{tian2020empirical} The RMST-based test enjoys the merit of easy interpretation and is distribution free.\cite{roystonRestrictedMeanSurvival2013} 
Moreover, it can be used to test superiority or non-inferiority. 

The test proposed by Liu et al. \cite{liuResamplingBasedTest2020} aims to detect crossing survival curves based on the area between the curves (ABC). It can capture the alternative of two crossing survival functions that produce the same RMST. The test obtains its p-value by (group-wise) bootstrapping, which allows different censoring distributions between groups. This test is shown to be more powerful than other distance-based tests such as
the modified Kolmogorov-Smirnov test \cite{fleming1980modified} and 
the generalized Cram{\'e}r-von Mises test \cite{schumacher1984two}. Since the test statistic quantifies the difference in absolute value, it cannot be used for superiority or non-inferiority testing.

%% file: 3_Simulations.tex
\section*{Simulation study}

To evaluate the performance of the presented methods, we employed extensive Monte Carlo simulations for different scenarios and settings. We simulated data for two groups under exponential, Weibull, Gompertz and log-normal distributions. Thus, we follow well-established recommendations on the choice of survival distributions for simulation studies. \cite{benderGeneratingSurvivalTimes2005}

\subsection*{Scenarios}
We considered four null scenarios, 
each with a different distribution function. For alternatives, we considered (i) four scenarios with proportional hazards, (ii) four scenarios with non-proportional and non-crossing hazards, and (iii) eight scenarios with crossing hazards. The concrete survival and hazard functions can be found in the Supplement, see Tables \ref{tab:Null}-\ref{tab:Cross2} therein. For each scenario we vary the group sizes (from $20$ to $100$), the censoring rates (from $0\%$ to $60\%$) and the censoring distributions (uniform, exponential) as listed in Table~\ref{tab:params} in the supplements. Thus, we studied $20$(scenarios)x$5$(sample sizes)x$4$(censoring rates)x$2$(censoring distributions) = $800$ different settings. We list three exemplary scenarios in Table~\ref{tab:3Exa}.

For each setting $5,000$ replications were performed. Throughout, we set the nominal size to be $0.05$. The actual type-I error and power were estimated by the rejection rates. For $10$ out of $800$ scenarios (all with small sample sizes) the MC test fails to provide a result in less than $0.5\%$ of the replications. In these cases, the mean rejection rate was thus computed for a slightly smaller number of observed results. Throughout, we used R~4.0.0 \cite{R} for all simulation.

\subsection*{Implementation details}

The LR as well as the PP can be called in R using the function \texttt{survdiff} from the \texttt{survival}\cite{survival-package} package. The concrete execution depends on the choice of rho (rho = 0 for LR and 1 for PP). 
The R package \texttt{TSHRC} \cite{TSHRC} contains the implementation of the TS test via the function \texttt{twostage}. 
The mdir is included in the R package \texttt{mdir.logrank} \cite{mdir.logrank}. 
Later, we refer to the test, mdir-x, where `x' stands for the number of weights considered. 
For the MC we use the weights proposed by Lin et al. \cite{lin2020alternative} and its implementation in the R package \texttt{nphsim} \cite{nphsim}. The KONP is implemented in the R package \texttt{KONPsurv} \cite{KONPsurv}. The packages provide tests based on two different test statistics, namely the log-likelihood ratio and the chi-squared test statistic. Since the authors did not detect any difference in performance we only consider the chi-squared test statistic (KONP).
An RMST-based test for two group comparisons is given in the R package \texttt{survRM2} \cite{survRM2}. The function used here is \texttt{rmst2}, where we need to define a truncation time tau. The published R code for the ABC test is provided on Github (https://github.com/LTTGH/RBT4TCSC). For both tests, $\tau$ was set to $90\%$ of the minimum of largest censored or uncensored event-time in two groups. \cite{tian2020empirical}

\begin{table}[H]
    \centering
    \begin{tabular}{|c|c c|}
    \hline
    Scenario & CDF & Survival and hazard curves \\
     \hline
     PH3 &  {$\!\begin{aligned}
              F_1(t) &= \exp(0.1)  \\
              F_2(t) &= \exp(1/28)  \end{aligned}$} & \raisebox{-.5\height}{
     \includegraphics[width=2.8in]{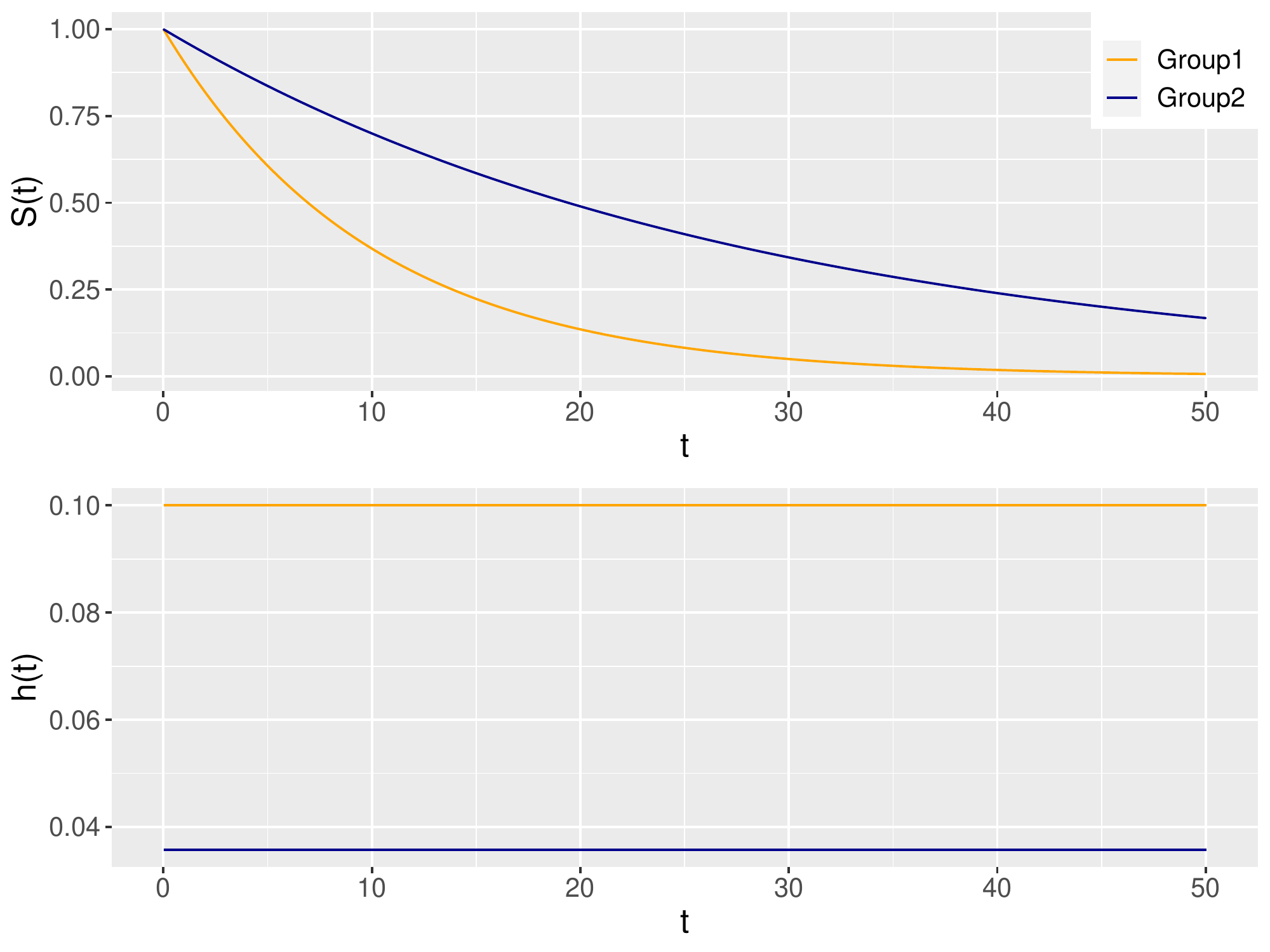}} \\
     \hline
     NPH4 &  {$\!\begin{aligned}
              F_1(t) &= \textrm{logN}(1.2,1.7)\\
              F_2(t) &= \textrm{logN}(2.4,1.3) \end{aligned}$} & \raisebox{-.5\height}{
     \includegraphics[width=2.8in]{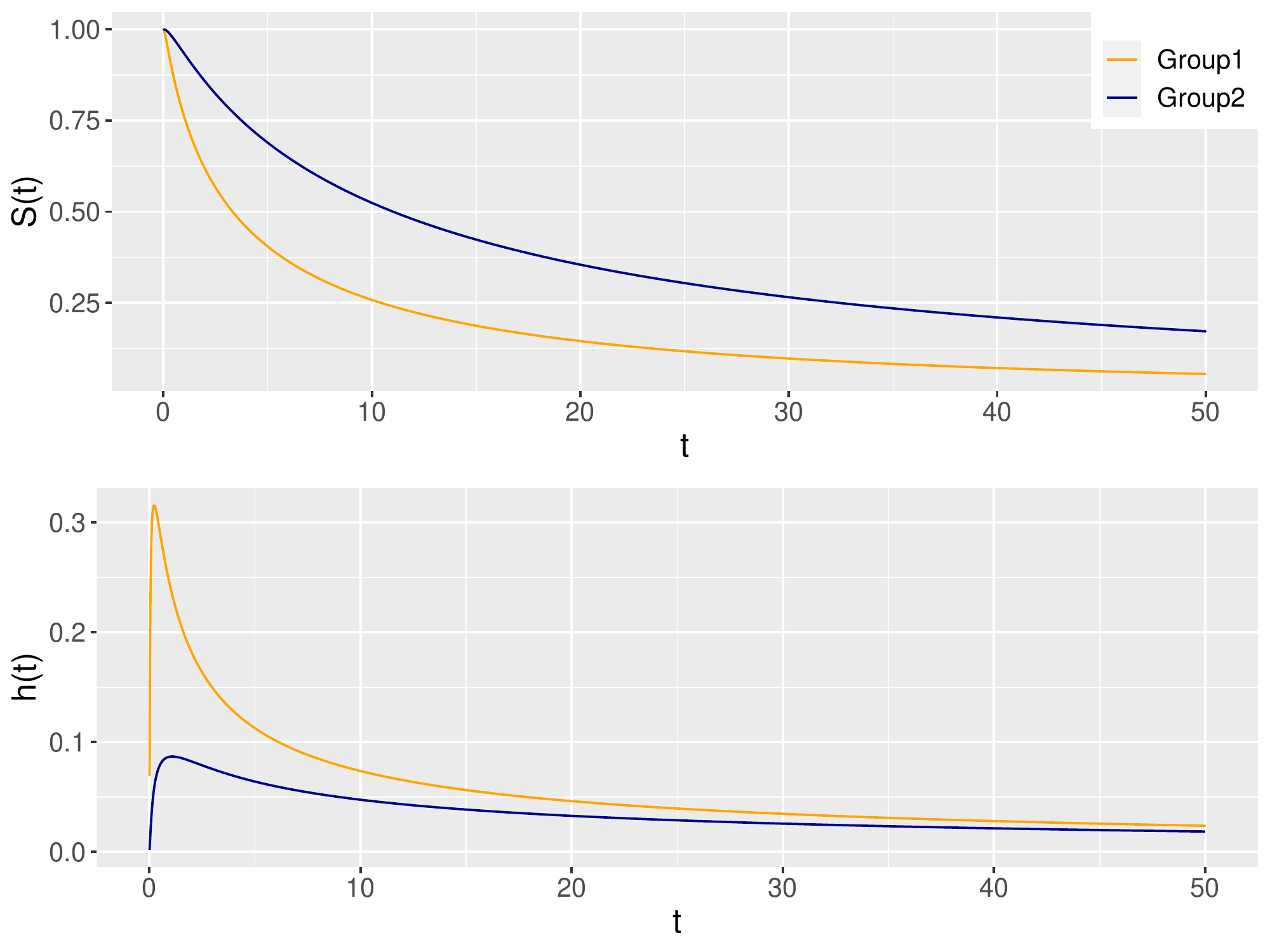}} \\
     \hline
     C3 \cite{liStatisticalInferenceMethods2015} &  {$\!\begin{aligned}
              F_1(t) &= \exp(1/12)\\
              F_2(t) &= \begin{cases}
                            \exp(0.25) & \text{$t \leq 2$}\\
                            \exp(1/35) & \text{$t > 2$}
                        \end{cases}
                \end{aligned}$} & \raisebox{-.5\height}{
     \includegraphics[width=2.8in]{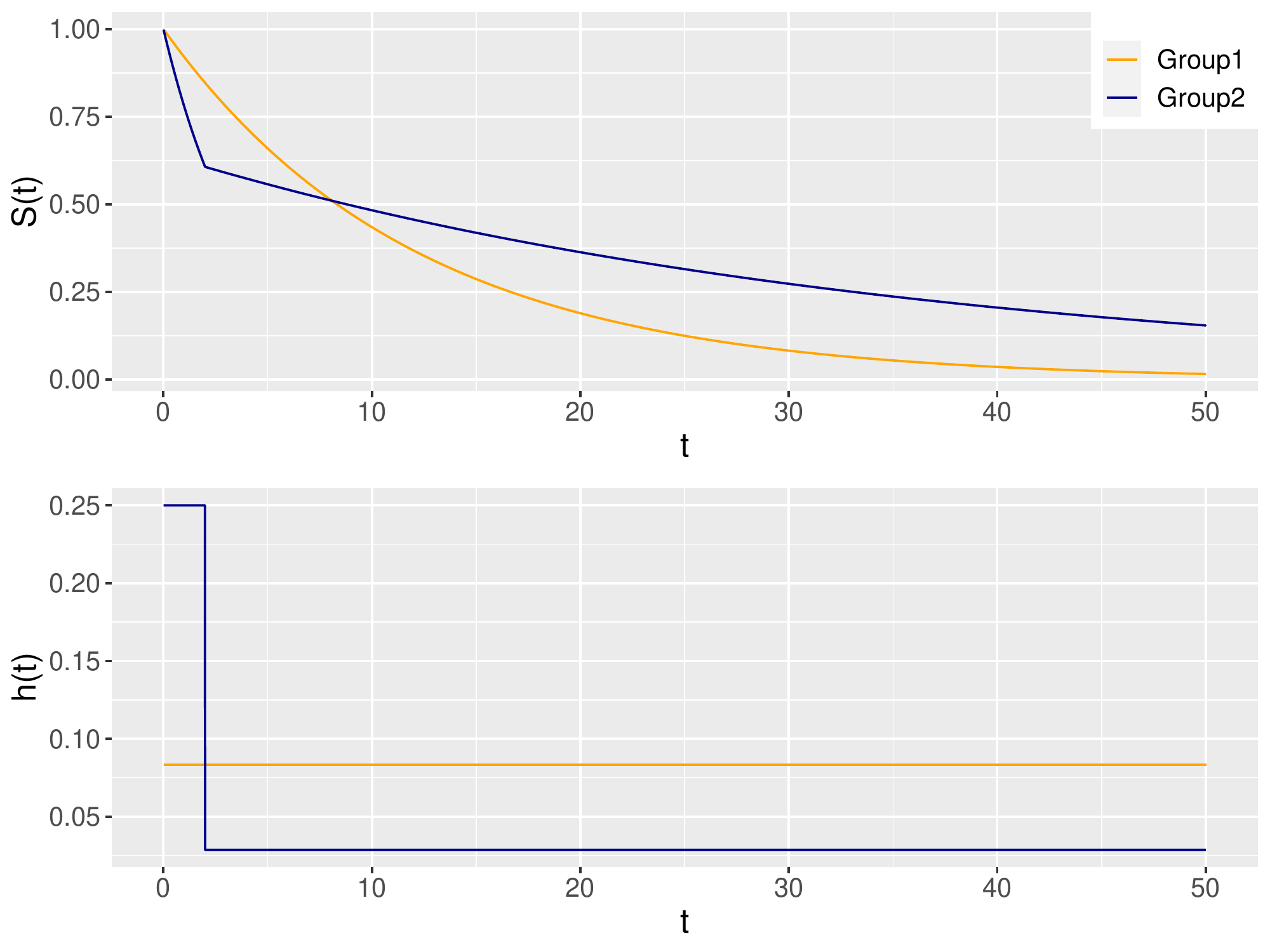}} \\
     \hline
    \end{tabular}
    \caption{Three exemplary scenarios. One scenario with proportional hazards (PH3), one with non-proportional and non-crossing hazards (NPH4) and one with crossing hazards (C3).}
    \label{tab:3Exa}
\end{table}

\subsection*{Results}

Here we only report the results under uniform censoring. The results under the exponential censoring are similar and provided in the Supplement. 

\begin{figure}[H]
    \centering
    \includegraphics[width=\textwidth]{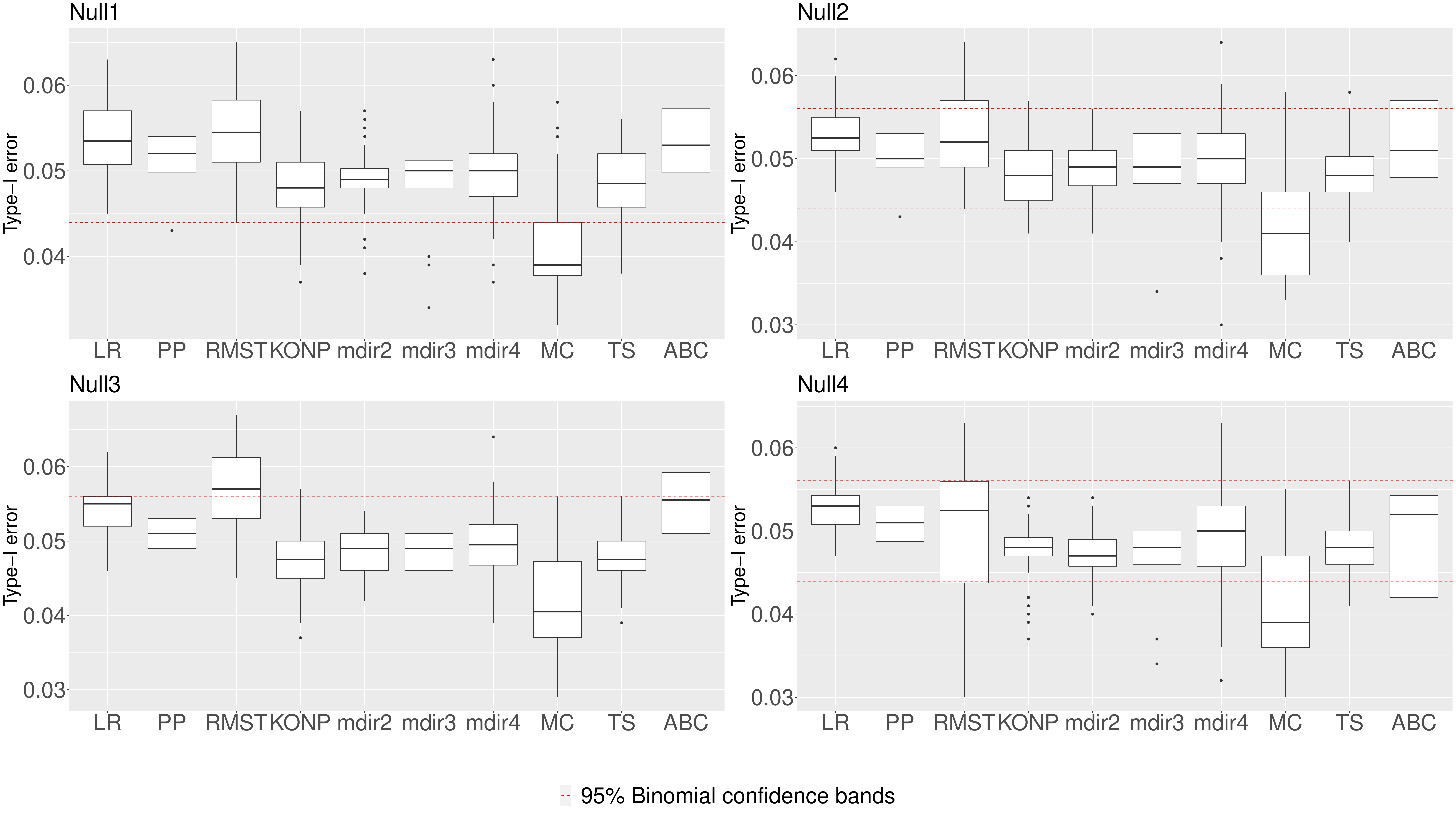}
    \caption{Type-I errors of the ten pairwise tests: LR log-rank test, PP peto-peto test, RMST restricted mean survival based test, KONP k-sample omnibus non-proportional hazards test, mdir2 mdir test with two weights, MC maxcombo test, TS two-stage test, ABC area between curves based test. The \textcolor{red}{red} dotted line represents the corresponding 95\% binomial interval [0.044, 0.056].} 
    \label{fig:BoxNull}
\end{figure}

\subsubsection*{Type-I-error}
Figure \ref{fig:BoxNull} compares the type I errors obtained by ten tests (nominal level $\alpha = 0.05)$ under the four considered scenarios. These employ different distributions that are commonly used in survival analysis \cite{benderGeneratingSurvivalTimes2005}. One boxplot summarizes $40$ data points (see Table \ref{tab:params} in the Supplement for the exact numbers) representing the size of the test for a specific parameter constellation based on $5,000$ simulation runs. The red-dotted lines display the binomial confidence intervals for the type-I error. For most of the tests it can be seen that the type-I error is usually within the red-dotted lines, implying a reasonable derivation from the significance level of $0.05$. However, the MC test is relatively conservative in all settings. Moreover, in the Null1 and Null3 scenario (top and bottom left) the RMST- and ABC-tests exhibit a rather liberal behavior. Nevertheless, all tests seem to control the type-I error reasonably well.

\subsubsection*{Power}
\begin{figure}[H]
    \centering
    \includegraphics[width=\textwidth]{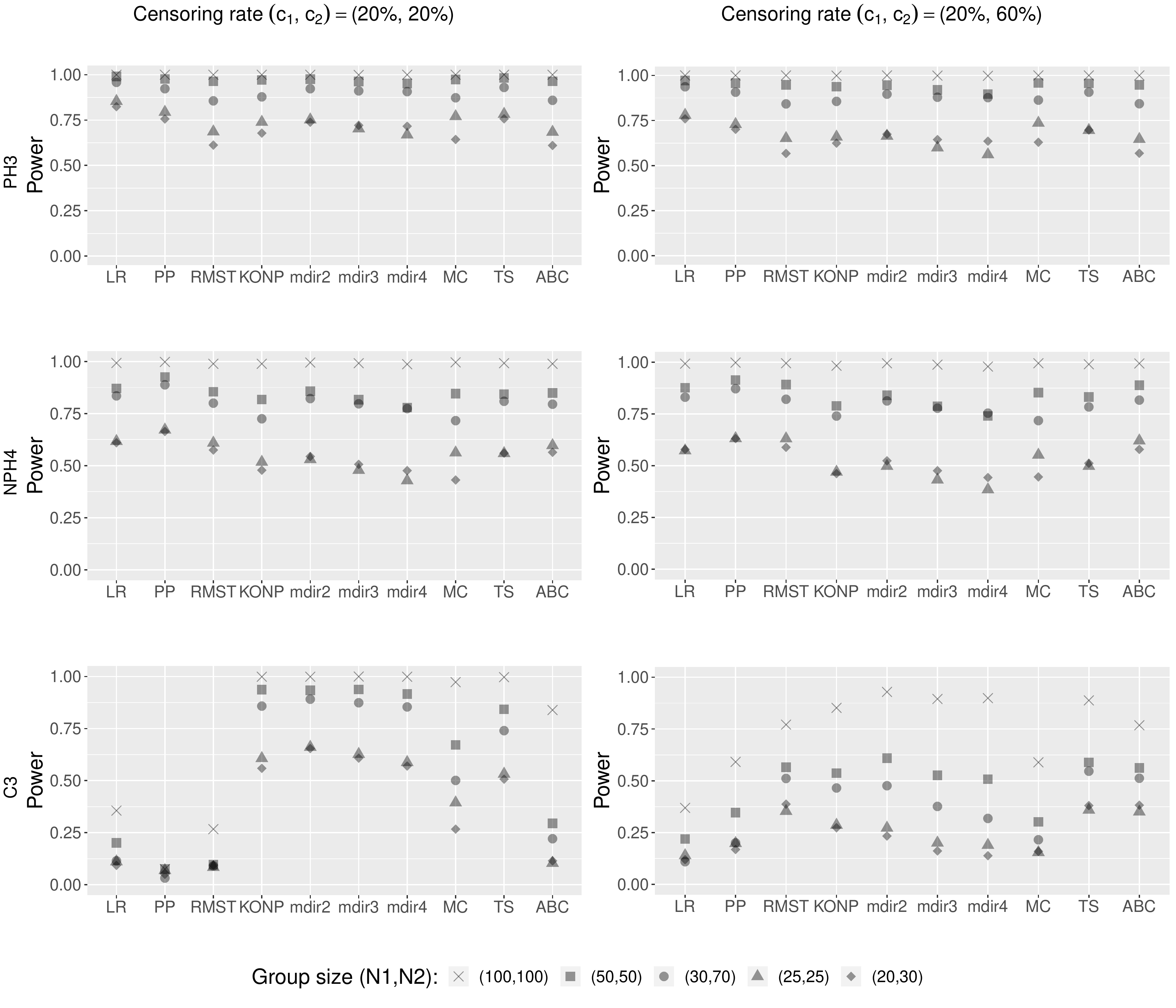}
    \caption{Power of the ten pairwise tests (nominal level $\alpha = 0.05)$ in representative scenarios under uniform censoring. PH3 proportional hazards scenario three, NPH4 non-proportional hazards scenario four, C3 crossing scenario three. LR log-rank test, PP peto-peto test, RMST restricted mean survival based test, KONP k-sample omnibus non-proportional hazards test, mdir2 mdir test with two weights, MC maxcombo test, TS two-stage test, ABC area between curves based test.}
    \label{fig:6er}
\end{figure}

We here summarize our findings for the $20$ different alternative scenarios. For ease of presentation, we only display three representative scenarios:

one for each category of possible relationships between hazards. Each row of Figure~\ref{fig:6er} represents a different scenario and the columns display different censoring rates.

\textbf{PH settings} The proportional settings are all generated using either the exponential or Weibull distribution. As expected, the LR test has the highest power over all settings and parameter combinations. In the first row of Figure~\ref{fig:6er} we observe that the other tests also provide good power for the different settings. However, with smaller sample sizes, they exhibit a noticeably power loss compared to the log-rank test. This effect is less pronounced for increased censoring. Similar patterns appear in all other scenarios with proportional hazards, see Figures \ref{fig:resProp1}-\ref{fig:resProp4} in the Supplement.

\textbf{Non-PH and non-crossing settings} Depending on the setting, the power of the tests varies significantly. In our example setting (see Table \ref{tab:3Exa}) depicted in the second row of Figure \ref{fig:6er}, the log-rank test has still high power for multiple parameter settings and also high censoring. At the same time, it does not dominate across all parameter combinations anymore. In the NPH4 scenario (second row in Table \ref{tab:3Exa}), the Peto-Peto test yields high power as the two survival functions have significant early difference. Analyzing Figures~\ref{fig:resNProp1}-\ref{fig:resNProp4}, the omnibus tests as well as the TS test are similarly powerful. Hence, under moderate violation of the PH assumption without crossing in hazard functions, the LR test can still be used while the omnibus is more robust against power loss. 

\textbf{Crossing hazards settings}

In the setting with crossing hazards (third row Figure \ref{fig:6er}), we observe a drastically lower power for the LR, the PP and the RMST based tests compared to all other methods under consideration. This pattern is present for low censoring as well as higher censoring. Nevertheless, high censoring leads to improvement in terms of power for some of the tests such as PP and RMST. A similar behavior can be observed for the ABC test: While the power for large groups drops slightly, the power derived from smaller data sets is higher than in the low censoring setting. The other tests under consideration lose power with high censoring regardless of the group sizes. Considering the other seven scenarios with crossing hazards available in the Supplement (Figures \ref{fig:resCross1}-\ref{fig:resCross8}), we see that in four of the eight scenarios the LR test is among the three tests with the lowest power over different censoring and sample size settings. The PP test is comparatively in seven out of eight scenarios and the RMST based test results in low power for all scenarios. However, the RMST based test often seems to have higher power for crossing scenarios with higher censoring. Consistently high power across the various scenarios is evident for the KONP test, the MC and the mdir tests. The TS test appears to be powerful for some scenarios but is less robust in terms of power than the omnibus tests. Finally, the ABC test has decent power for most of the scenarios but is no competitor for the omnibus tests and the TS test. Regarding the choice of weights for the mdir test, it can be seen that except for one crossing scenario the mdir2 test including the log-rank and crossing weight is as powerful as or more powerful than the mdir3 or mdir4. Hence, we would recommend using the mdir test with these two weights only.

In summary, it has to be further investigated why the MC test is so conservative (Figure \ref{fig:BoxNull}). A reasonable assumption is the small sample size in the groups. Taking the conservative behavior into account, we can assume that larger sample sizes might also lead to higher power in the alternative scenarios. The results of Lin et al. \cite{lin2020alternative} support this assumption. They considered much larger sample sizes starting from $300$ and did not observe a similar behavior. The results show that it is adequate to include two different weighted LR statistics in the mdir test. The RMST-based test cannot be recommended in situations with crossing hazards. Globally, we do recommend the use of omnibus tests such as MC, KONP or mdir when no prior knowledge is available. They show robust power behavior for proportional, non-proportional and crossing scenarios.

%% file: 4_Data_Example.tex
\section*{Real Data Example}

We evaluate the performance of the considered tests on real data from a clinical trial for $131$ elderly patients with advanced liposarcoma or leiomyosarcoma, where the overall survival functions under two treatments (Dacarbazine and Trabectedin) show a clear cross at a late time period. \cite{jonesEfficacyTolerabilityTrabectedin2018}

\begin{figure}[H]
    \centering
    \includegraphics[width = 0.8\textwidth]{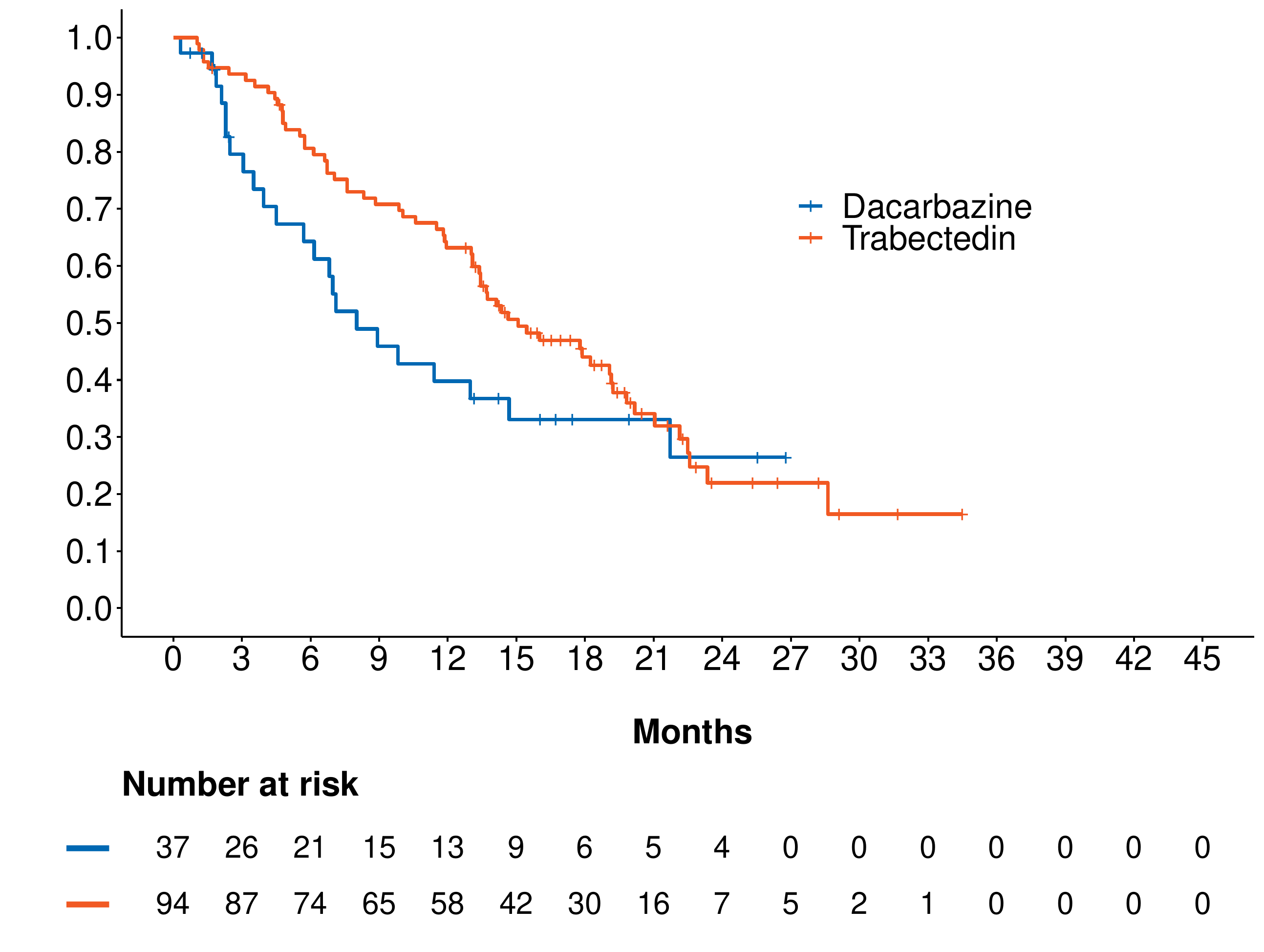}
    \caption{Reconstructed Kaplan-Meier curves of overall survival in elderly patients with sarcoma from Jones at al. \cite{jonesEfficacyTolerabilityTrabectedin2018}}
    \label{fig:Jones}
\end{figure}

As in Dormuth et al. \cite{dormuth2022}, we reconstructed the patient-level data using the state-of-the-art reconstruction algorithm by Guyot et al. \cite{guyotEnhancedSecondaryAnalysis2012}. The Kaplan-Meier plot using the reconstructed data is shown in Figure~\ref{fig:Jones}. It is apparent that the Kaplan-Meier curves depart from each other early but converge at later times. The quality of the reconstructed data is examined to be sufficiently satisfactory (See Table~\ref{tab:Jones} in the Supplement).

The p-values for eight considered tests are listed in Table~\ref{tab:results_Jones}. All the considered methods except LR and MC succeed to reject the null hypothesis at significance level 0.05 with mdir2 giving the strongest note.

\begin{table}[H]
    \centering
    \footnotesize
    \begin{tabular}{|cccccccccc|}
    \hline
         LR & PP & RMST & KONP &  mdir2 & mdir3 & mdir4& MC & 2ST & ABC\\
         \hline
         0.17 &  \textbf{0.03}  & \textbf{0.02} &    \textbf{0.04 }& \textbf{0.01 }& \textbf{0.03}  & 0.05 & 0.06 & \textbf{0.05}  & \textbf{0.02}
         \\
         \hline
    \end{tabular}
    \caption{P-values of the ten pairwise tests (nominal level $\alpha = 0.05)$ applied to the reconstructed data from Jones et al. \cite{jonesEfficacyTolerabilityTrabectedin2018}. LR log-rank test, PP peto-peto test, RMST restricted mean survival based test, KONP k-sample omnibus non-proportional hazards test, mdir2 mdir test with two weights, MC maxcombo test, TS two-stage test, ABC area between curves based test.}
    \label{tab:results_Jones}
\end{table}

%% file: 5_Discussion.tex
\section*{Discussion}
We investigated the type-I error and power of the gold-standard log-rank test and various recent tests that are recommended as alternatives in case of potential proportional hazards violation. To this end we conducted an extensive simulation study including $20$ representative scenarios. In the null settings most  tests respected the type-I error. Only the MC test appears to be more conservative for small sample sizes. Regarding power, the simulation study indicates that 
the log-rank test does not experience a drastic loss compared to the other methods in case of non-proportional and non-crossing scenarios. However, if crossings are present, the power difference among the tests is much more pronounced with good performances for all omnibus tests. In fact, especially the 
 KONP and mdir omnibus tests show a more stable power over all scenarios. Regarding the mdir test, no advantage in inclusion of more weights for the mdir test was found in our settings. We therefore recommend using the default setting of the test if no prior knowledge is available. 
 
 {\bf Limitations of our study.} We only investigated the methods' performance for small to moderate sample sizes. 
 In this paper we only discussed the two-sided testing problem since not all procedures have existing versions for testing superiority or  non-inferiority. Here, we see further research potential as this kind of one-sided testing versions only exist for the weighted LR, the RMST-based and the mdir tests. \cite{ananthakrishnanCriticalReviewOncology2021, ditzhaus2019wild} A similar statement holds for $k$-sample or more general ANOVA settings. \cite{gorfineKsampleOmnibusNonproportional2019, ditzhaus2021casanova}  We therefore recommend to also investigate and compare their mathematical properties, e.g. relative efficiencies.
 
 In order for the recommended procedures to find their way into biostatistical practice, methods for accurate sample size calculation are needed. Otherwise, inclusion in study protocols as well as the ability to draw sufficiently powered conclusions is not given.  Furthermore, statistical significance alone does not always corroborate clinical relevance. Statistical measures of treatment effect such as confidence interval for meaningful estimands/parameters are necessary to improve the tests' interpretability. Furthermore, we recommend to also investigate and compare their mathematical properties, e.g. relative efficiencies.

%% file: 6_Conclusion.tex
\section*{Acknowledgments}
This work has been partly supported by the Research Center Trustworthy Data Science and Security (https://rc-trust.ai), one of the Research Alliance centers within the https://uaruhr.de. 
The authors gratefully acknowledge the computing time provided on the Linux HPC cluster at Technical University Dortmund (LiDO3), partially funded in the course of the Large-Scale Equipment Initiative by the German Research Foundation (DFG) as project 271512359.

%% file: 7_Appendix.tex
\appendix

\setcounter{table}{0}   
\renewcommand{\thetable}{S\arabic{table}}
\setcounter{figure}{0}   
\renewcommand{\thefigure}{S\arabic{figure}}
\setcounter{equation}{0}   
\renewcommand{\theequation}{S\arabic{equation}}

\section*{Methods}
\subsection*{Weighted Log-rank-Tests}
The standard to compare two Kaplan-Meier curves of interest is the log-rank test (LR).\cite{singh2011survival} It belongs to a class of weighted tests for equality of survival that use the difference between the expected and observed number of deaths to obtain a test statistic. \cite{kleinSurvivalAnalysisTechniques2003} Defining for the different tests is the choice of the weight function in the test statistic:

\begin{equation}\label{eqn:def_logrank}
     Z_n(w) = \frac{\sum_{i=1}^D w_i (N_{i1}- Y_{i1} \frac{N_i}{Y_i})}{\sqrt{\sum_{i=1}^D w_i^2 \frac{Y_{i1}}{Y_i}(1-\frac{Y_{i1}}{Y_i}) \frac{Y_i - N_i}{Y_i -1}N_i}}.
 \end{equation}

Besides the weight function $w_i$ we consider the number of events $N_{ij}$ in group $j=1,2$ at event time $t_i, ~ i = 1,...,D$ and the corresponding number at risk $Y_{ij}$.\cite{kleinSurvivalAnalysisTechniques2003} The corresponding number at risk and number of events in the pooled sample are denoted as $Y_i$ and $N_i$ respectively. To obtain the log-rank test statistic we set $w_i \equiv 1$. This test has optimal power in situations where the hazard rates are proportional to each other, but loses power if this assumption does not hold. \cite{kleinSurvivalAnalysisTechniques2003} Under the null hypothesis of equal survival, the test statistic for two groups is approximately $\chi^2_1$ distributed. \cite{kleinSurvivalAnalysisTechniques2003}

Since the log-rank test is often applied, it is implemented in various statistical software such as the free open-source software R. There, it can be conducted using the function survdiff from the survival\cite{survival-package} package. 

\subsection*{Peto-Peto test}
Another test belonging to the same group as the log-rank test is the Peto-Peto test. As implied above, the Peto-Peto test is characterized by a different weight function than the log-rank test. Setting the weights in \eqref{eqn:def_logrank} to $w_i \equiv \Tilde{S}(t_i)$, we obtain a test that is more sensitive to early differences in survival. With $\Tilde{S}(t_i)$ an estimate of the pooled survival function at event time $t_i, ~ i = 1,...,D$. \cite{kleinSurvivalAnalysisTechniques2003} It can be conducted using the survdiff function in R as well. \cite{survival-package} One challenge of using weighted log-rank tests is the choice of weights, since the behavior of the hazard rates is usually unknown. 

\subsection*{mdir test}

Approaches allowing the combination of multiple weights provide a flexible solution without the need of prior information. \cite{BrendelETAL2014,DitzhausFriedrich2018,ditzhaus2019wild} Ditzhaus and Friedrich \cite{DitzhausFriedrich2018} revisited such a test for multiple alternatives and introduced a permutated version of the test allowing to handle small sample sizes. They also created an R package \textit{mdir.logrank} \cite{mdir.logrank}.The test assumes equality of survival under the null hypothesis. The alternative(s) depend on the choice of weights, as default in R a combination of the log-rank weight $w_{i}^{(1)}\equiv 1$ and a crossing weight $w_{i}^{(2)} = 1 - 2\widehat{S}(t_i)$ is implemented. With two weights, the test statistic has the following studentized quadratic form
 \[S_n=(Z_n(w^{(1)}), Z_n(w^{(2)}))\,\hat{\Sigma}^-_n\, (Z_n(w^{(1)}), Z_n(w^{(2)}))^T.\]
Here, $ \hat{\Sigma}^-_n$ denotes the Moore-Penrose inverse of the empirical covariance matrix of $(Z_n(w^{(1)}),Z_n(w^{(2)}))$. Then, the test statistic follows a $\chi^2$-distribution under $H_0$ \cite{DitzhausFriedrich2018}. It is possible to combine more weights in order to cover more alternatives if needed. For a combination of $m$ weighted log-rank tests, the test statistic is defined as 
\[S_n=(Z_n(w^{(1)}),..., Z_n(w^{(m)}))\,\hat{\Sigma}_n\, (Z_n(w^{(1)}),..., Z_n(w^{(m)}))^T,\]

with the entries of $\hat{\Sigma}_n$ being
\[(\hat{\Sigma}_n)_{r,s}= \frac{n}{n_1n_2} \int_{[0,\infty)}w_s(\hat{F}(t-))w_r(\hat{F}(t-))\frac{Y_1(t)Y_2(t)}{Y(t)}d\hat{A}(t), ~~~ (r,s=1,...,m), \]
with $\hat{A}$ the Nelson-Aalen estimator. 

\subsection*{Maxcombo}
Another way to combine multiple weighted log-rank tests is the MaxCombo test. The test statistic is the maximum over standardized weighted log-rank tests, where the weight function is of the Flemming-Harrington type \cite{lee2007versatility}. Thus, the test statistic can be given as \[Z_\text{max} = \text{max}_{\rho,\gamma}(Z_{FH(\rho, \gamma)})\] with $Z_{FH(\rho, \gamma)}$ the standardized weighted log-rank test statistics. We used the weights proposed by Lin et al. \cite{lin2020alternative}: FH(0,0), FH(0,1), FH(1,1) and FH(1,0). Critical values of the MaxCombo test can be derived based on the asymptotic normality of $Z$ \cite{yang2005combining}. It is implemented in the nphsim \cite{nphsim} package in R.

\subsection*{Two-stage test}
The two-stage procedure introduced by Qiu and Sheng\cite{qiu2008} provides a test with reliable power in cases of PH and non-PH. The procedure got its name from the sequential testing approach, where a test for proportional hazards such as any log-rank type test is performed in a first step. Then, only if the test is not able to reject the null hypothesis an asymptotically independent test specifically for crossing is conducted. Therefore, the authors developed a new weighting scheme with the main concept of changing signs before and after possible crossing points. The new specific weights for the weighted log-rank test are  $w_i^{(m)}=-1$ for all $i\leq m$ and $w_i^{(m)}=c_m$ for $i>m$ with $m$ the potential crossing times. We compute the positive constant $c_m$ such that the second test is asymptotically independent of the first test. Combining this, we obtain the overall test statistic as the supremum of the weighted log-rank tests with $D_\epsilon \leq m \leq D - D_\epsilon$:
\begin{align*}
     V &= \sup_{D_\epsilon \leq m \leq D-D_\epsilon} Z_n(w^{(m)}),
 \end{align*}
 where $\epsilon>0$ small and $D_\epsilon$ equals the integer part of $D \cdot \epsilon$.

The independence between the two stages allows to calculate the corresponding p-value using:

 \begin{align*}
p = \left\{
 \begin{matrix}
 p_1&, ~ \text{ if } p_1 \leq \alpha_1\\
 \alpha_1 + p_2(1-\alpha_1)&, \text{ otherwise,}
 \end{matrix}
 \right.
 \end{align*}
 
 with $\alpha_1 = \alpha_2 = 1- \sqrt{1-\alpha}$. Thus, the sequential testing is taken into account and the control of the type-I error is ensured. $p_1$ the p-value of the first test can be obtained using the standard normal distribution while $p2$ is obtained via bootstrap.To conclude, the two stage test rejects the null hypothesis if the first test rejects it or if it does not the second one does. The authors emphasize that the test has higher power with fewer crossings and if the number of crossings is known. The test can be conducted in R using the function twostage from the package TSHRC \cite{TSHRC}.

\subsubsection*{RMST-based test}
Often other effect measures are considered to compare the survival in two groups without emphasizing the different null hypotheses they consider. One example is the restricted mean survival (RMST). In order to avoid the problem of crossing hazards once and for all, RMST based tests have been introduced.\cite{kim2017restricted} Since they do not use hazards, they do not require proportional hazards. Still they test a slightly different null hypothesis. Instead of testing for equality in survival they test for equality in RMST to a pre-defined time $\tau$. The RMST can be interpreted as the mean event-free survival time up to $\tau$. Tian et al.\cite{tianEfficiencyTwoSample2018} propose to estimate the group difference in RMST by the difference of the areas under the two Kaplan-Meier (KM) curves up to $\tau$ :
\[\widehat{\Delta}(\tau) = \int^\tau_0 \widehat{S}_1(u) - \widehat{S}_2(u)du.\] 

\subsubsection*{ABC-based test}
Unlike RMST-based tests, a test based on the area between the curves (ABC) is actually designed to detect crossing survival curve departures. In fact, it is possible to observe two crossing survival curves with the same RMST. Liu et al. \cite{liuResamplingBasedTest2020} defined the area between the curves as
\[T_n = \sqrt{n} \int_0^\tau |\hat{S}_1(u) - \hat{S}_2(u)|du\]
for a time window $[0,\tau]$ of interest. The corresponding p-value is calculated using a group-wise bootstrap, which allows for the survival data to have different censoring distributions. The method is not available as an R package yet, but the code can be found in the supplements.

\subsubsection*{KONP test}
Tests based on sample-space partition such as presented by Heller et al. \cite{heller2016consistent} can be applied to k-sample problems.\cite{gorfineKsampleOmnibusNonproportional2019} So far, they are often employed for uncensored data and Gorfine et al.\cite{gorfineKsampleOmnibusNonproportional2019} introduce a test based on this concept that also tackles right censored data. Here, we consider only the 2-sample case.

For non-censored data, 2 random samples $A_1$ and $A_2$ are drawn from the distributions $F_1$ and $F_2$. For an $X_i$ in the sample $A_i$ and an $X_j$ from one of the random samples. The partition is then based on the tuple $(i,j)$, resulting four different groups depending on the absolute distance between the other observations to $X_i$ and $X_i$ to $X_j$. Exemplary, one group contains the number of observations that are in $A_i$ and their absolute distance to $X_i$ is smaller than $|X_i - X_j|$.Based on the number of observations in the four resulting groups a 2x2 contingency table is created. For censored data each of the $K$ samples includes information about the time (either censoring or event), the censoring indicator and the group.\cite{gorfineKsampleOmnibusNonproportional2019}

For right-censored data, the Kaplan-Meier curves are used instead of the empirical distribution function to draw random samples. For the estimation of the contingency table, only observed events are considered. Furthermore, only observations before a defined maximum time are included. The contingency tables are summarized using either the chi-squared test statistic (KONP\_chi) or the log-likelihood ratio statistic (KONP\_llr). These are averaged for all the resulting tables for all pairs $(i,j)$ in a test statistic Q. The authors emphasize that the number of tables in the case of right-censored data depends on the data and especially on the censoring pattern.\cite{gorfineKsampleOmnibusNonproportional2019}

To obtain the p-values a permutation procedure is introduced. Instead of assigning new labels $Y_i$ to each observation, an imputation step is additionally performed. If the permuted group is different than the original group, a new event-time is generated based on the KM-estimator of the permuted group. The authors point out that this imputation leads to additional variability in the p-value. The number of permutations should thus be increased by the number of imputations performed. \cite{gorfineKsampleOmnibusNonproportional2019} The test for both aforementioned summary statistics is implemented in the R package KONPsurv \cite{KONPsurv}.

\newpage
\section*{Simulation Scenarios} 


\begin{table}[ht]
    \centering
    \begin{tabular}{|c|c|c|c|}
        \hline
        N1 & N2 & C\_rate1 & C\_rate2 \\
        \hline
        100 & 100 & 0 & 0\\
        50  & 50 & 0 & 0\\
        30	& 70	& 0	& 0\\
        25	& 25	& 0	& 0\\
        20	& 30	& 0	& 0\\
        100	& 100	& 0.2	& 0.2\\
        50	& 50	& 0.2	& 0.2\\
        30	& 70	& 0.2	& 0.2\\
        25	& 25	& 0.2	& 0.2\\
        20	& 30	& 0.2	& 0.2\\
        100	& 100	& 0.2	& 0.4\\
        50	& 50	& 0.2	& 0.4\\
        30	& 70	& 0.2	& 0.4\\
        25	& 25	& 0.2	& 0.4\\
        20	& 30& 	0.2	& 0.4\\
        100	& 100	& 0.2	& 0.6\\
        50	& 50	& 0.2	& 0.6\\
        30	& 70	& 0.2	& 0.6\\
        25	& 25	& 0.2	& 0.6\\
        20	& 30	& 0.2	& 0.6\\

        \hline
    \end{tabular}
    \caption{List of parameter combinations for the 20 different settings per censoring distribution. Leaving a total of 40 parameter combinations for each scenario considering uniform distribution and exponential distribution for censoring.}
    \label{tab:params}
\end{table}

\newgeometry{top=1.5cm}

\begin{table}[H]
    \centering
    \begin{tabular}{|c|c c|}
    \hline
    Scenario & CDF & Visualization of the survival and hazard curves \\
    \hline
         Null1 &  $F_1(t) = F_2(t) = \textit{Weibull}(1.5, 30) $ & \raisebox{-.5\height}{
      \includegraphics[width=2.8in]{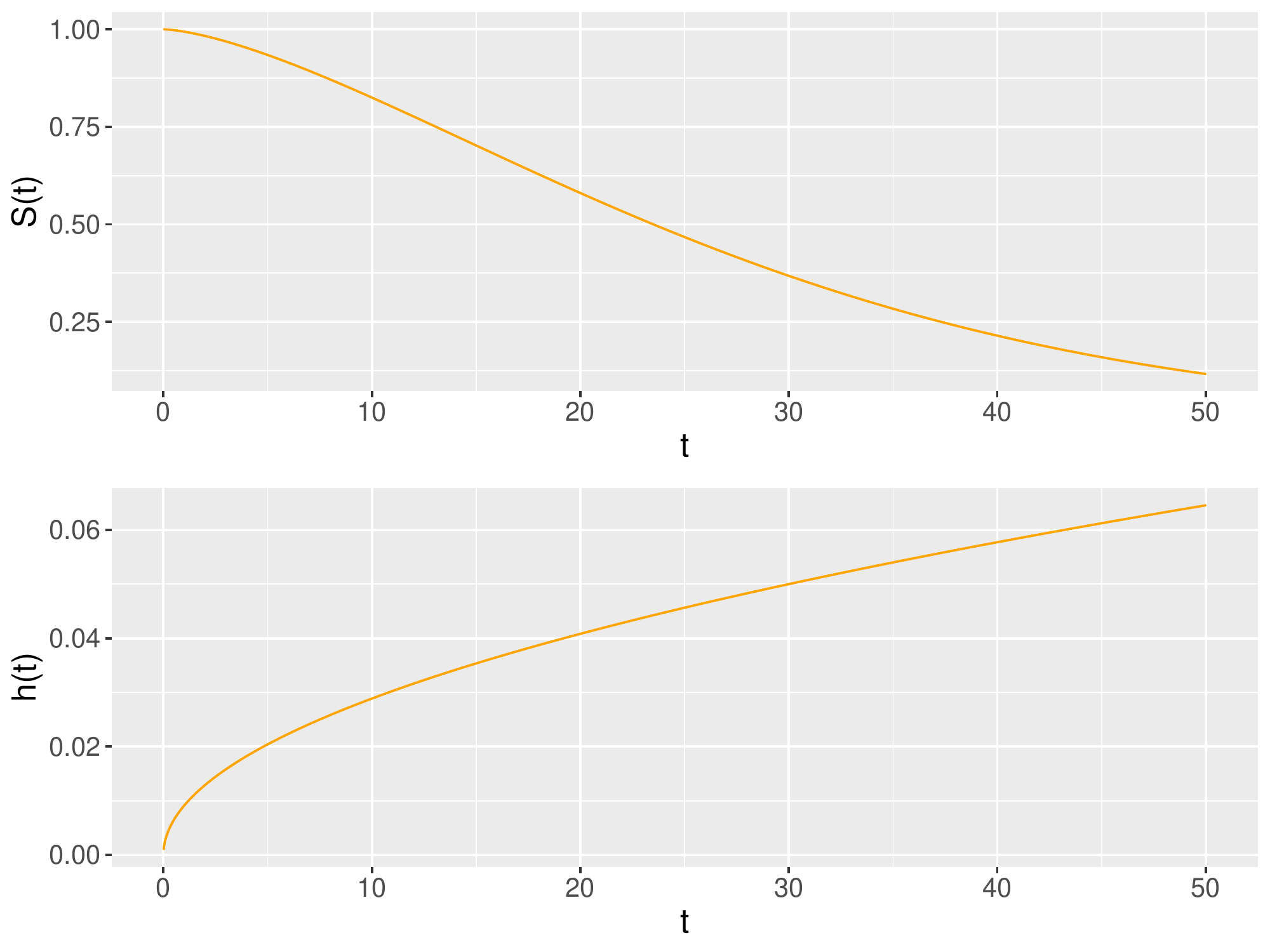}} \\
      \hline
      Null2 &  $F_1(t) = F_2(t) = \textit{Exp}(0.1) $ & \raisebox{-.5\height}{
      \includegraphics[width=2.8in]{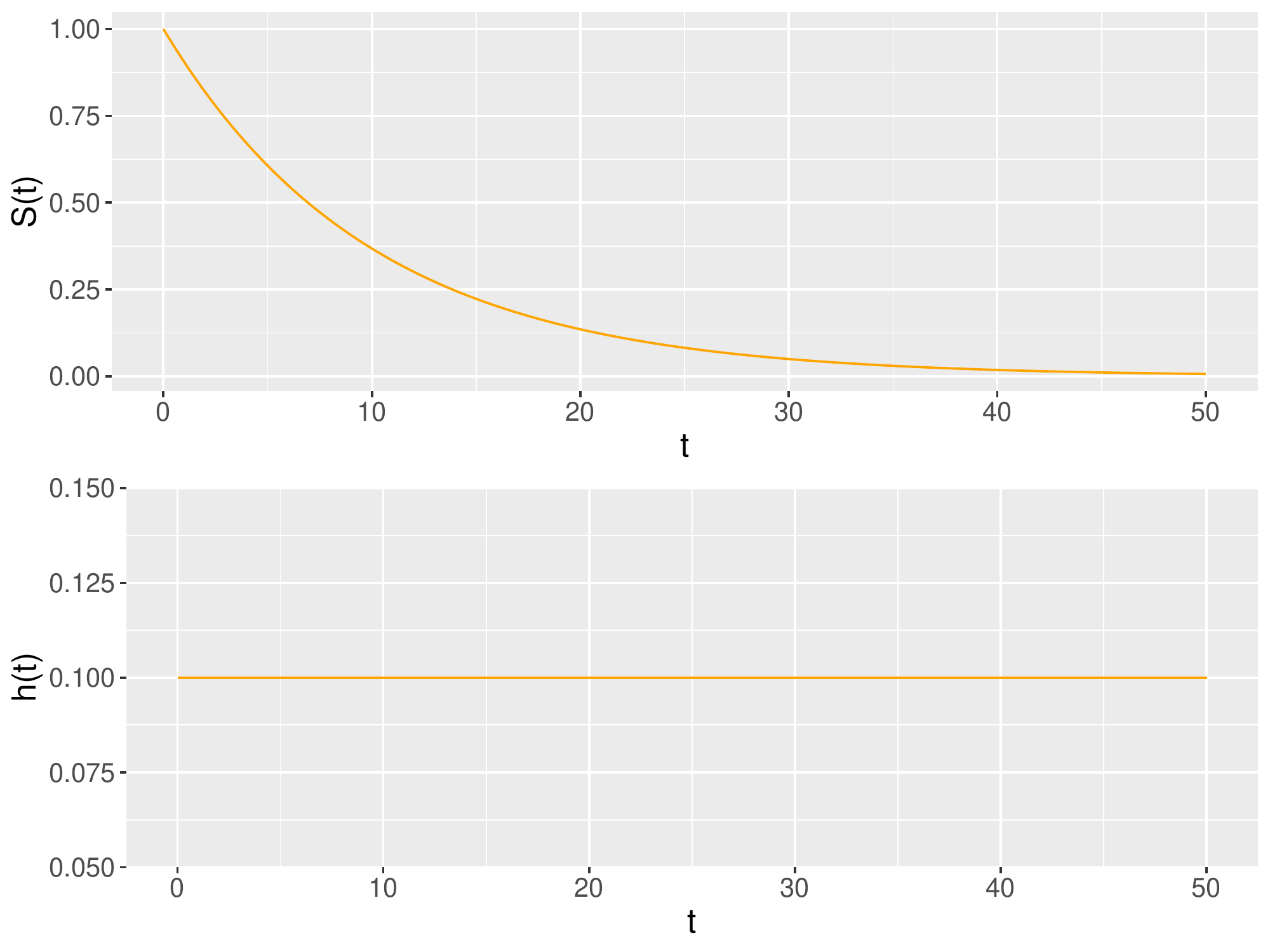}} \\
      \hline
      Null3 &  $F_1(t) = F_2(t) = \textit{Gompertz}(0.2, 0.4) $ & \raisebox{-.5\height}{
      \includegraphics[width=2.8in]{Plots/NullExp.pdf}} \\
      \hline
      Null4 &  $F_1(t) = F_2(t) = \textit{Lognormal}(1.2, 1.7) $ & \raisebox{-.5\height}{
      \includegraphics[width=2.8in]{Plots/NullExp.pdf}} \\
      \hline
    \end{tabular}
    \caption{Scenarios with equal hazard functions.}
    \label{tab:Null}
\end{table}

\begin{table}[H]
    \centering
    \begin{tabular}{|c|c c|}
    \hline
    Scenario & CDF & Visualization of the survival and hazard curves \\
    \hline
         PH1 &  {$\!\begin{aligned} 
               F_1(t) &= \textit{Weibull}(0.6, 8)  \\   
               F_2(t) &= \textit{Weibull}(0.6, 4)  \end{aligned}$} & \raisebox{-.5\height}{
      \includegraphics[width=2.8in]{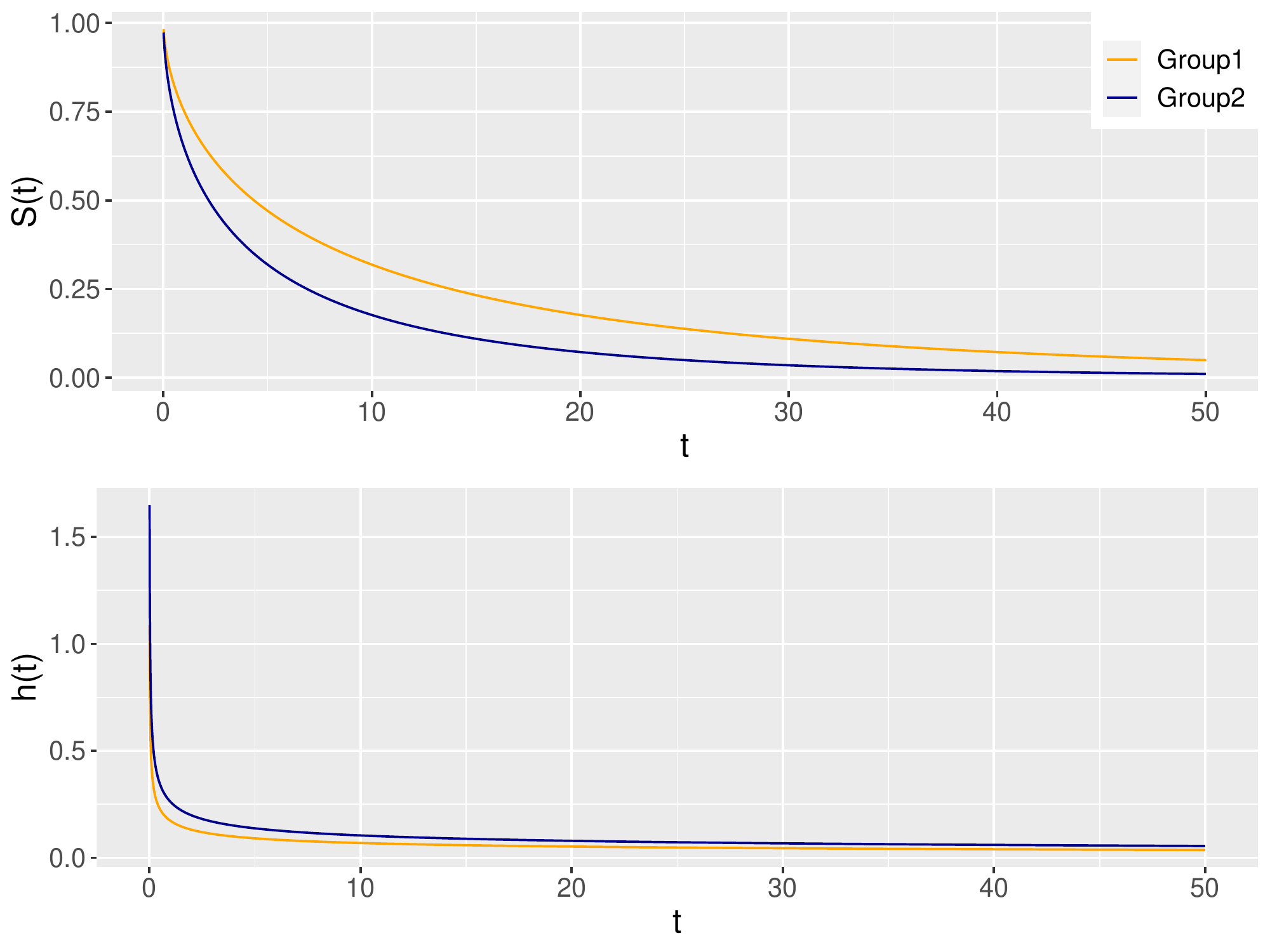}} \\
      \hline
      PH2 &  {$\!\begin{aligned} 
               F_1(t) &= \textit{Weibull}(1.3, 8)  \\   
               F_2(t) &= \textit{Weibull}(1.3, 4)  \end{aligned}$} & \raisebox{-.5\height}{
        \includegraphics[width=2.8in]{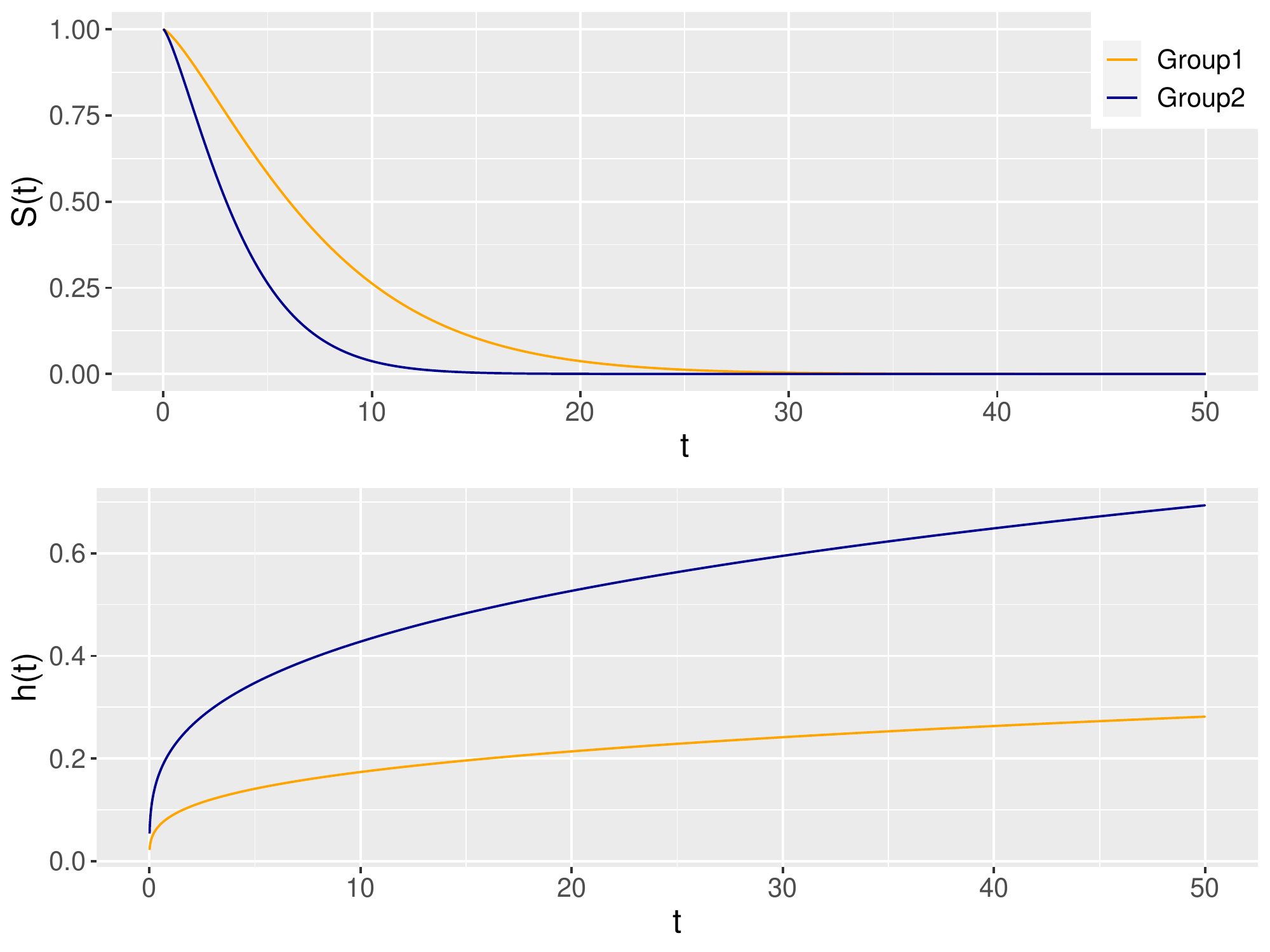}} \\
      \hline
      PH3 &  {$\!\begin{aligned} 
               F_1(t) &= \textit{Exp}(0.1)  \\   
               F_2(t) &= \textit{Exp}(1/28)  \end{aligned}$} & \raisebox{-.5\height}{
      \includegraphics[width=2.8in]{Plots/ExpProp1.pdf}} \\
      \hline
      PH4 \cite{liStatisticalInferenceMethods2015} &  {$\!\begin{aligned} 
               F_1(t) &= \textit{Exp}(0.5)   \\   
               F_2(t) &= \textit{Exp}(0.2) \end{aligned}$} & \raisebox{-.5\height}{
      \includegraphics[width=2.8in]{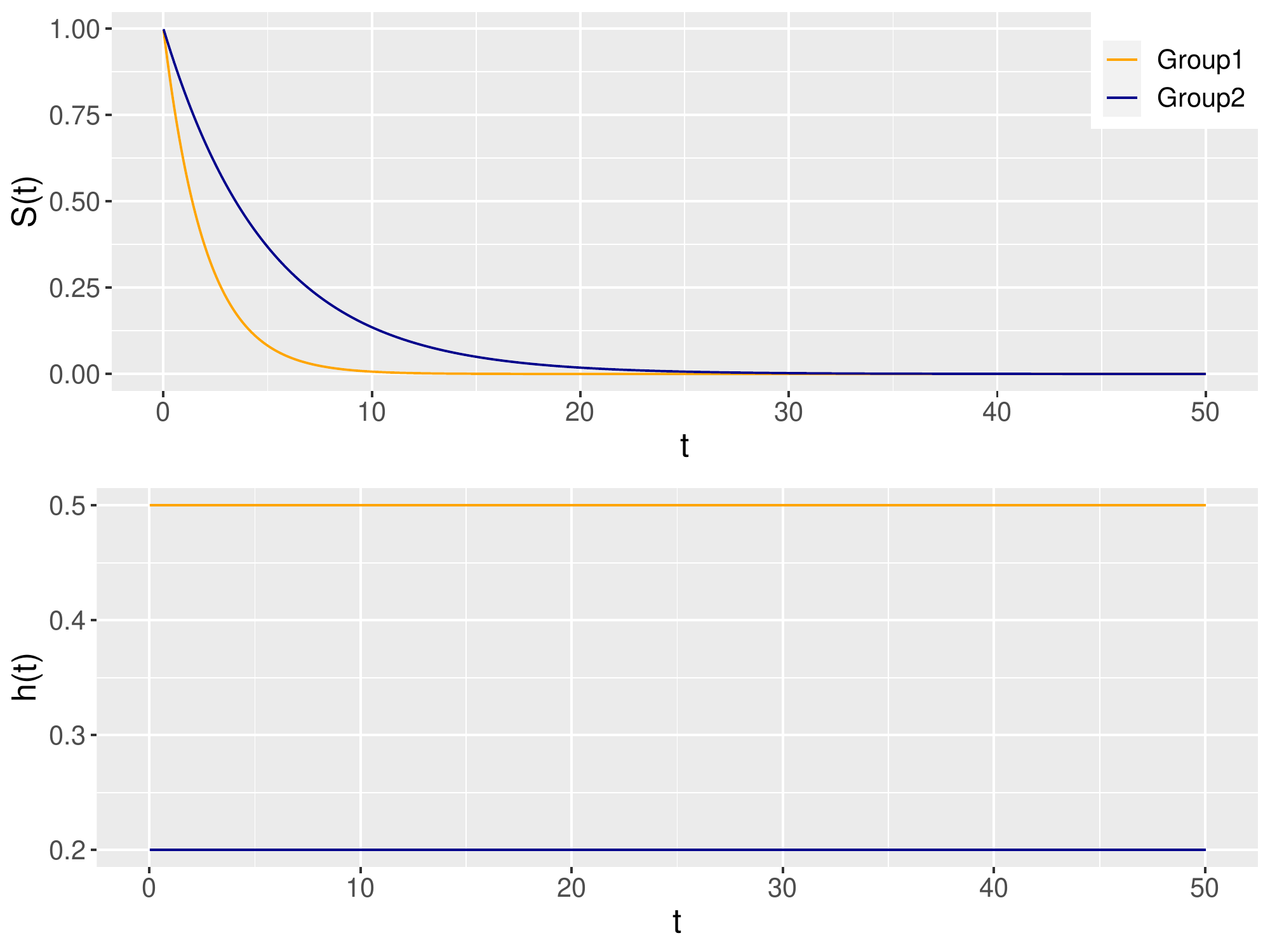}} \\
      \hline
    \end{tabular}
    \caption{Scenarios with proportional hazard functions.}
    \label{tab:Prop}
\end{table}

\begin{table}[H]
    \centering
    \begin{tabular}{|c|c c|}
    \hline
    Scenario & CDF & Visualization of the survival and hazard curves \\
    \hline
         NPH1  &  {$\!\begin{aligned} 
               F_1(t) &= \textit{Weibull}(2.5, 30)  \\   
               F_2(t) &= \textit{Weibull}(3, 25)  \end{aligned}$} & \raisebox{-.5\height}{
      \includegraphics[width=2.8in]{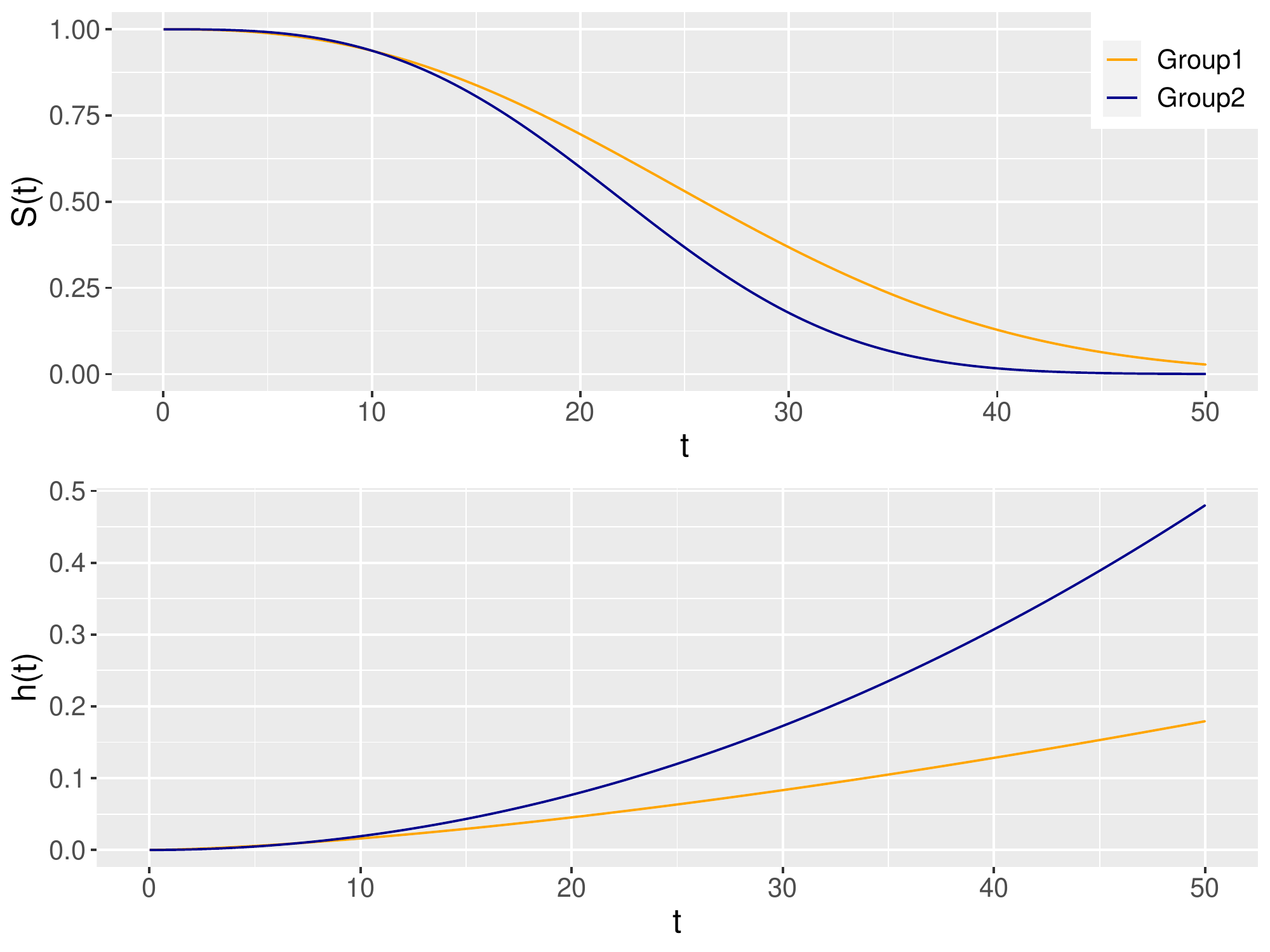}} \\
      \hline
      NPH2 \cite{gorfineKsampleOmnibusNonproportional2019} &  {$\!\begin{aligned} 
               F_1(t) &= \textit{Exp}(1)\\
               F_2(t) &= \begin{cases}
                            \textit{Exp}(1) & \text{$t \leq 0.3$}\\
                            \textit{Exp}(2) & \text{$t > 0.3$}
                        \end{cases} \end{aligned}$} & \raisebox{-.5\height}{
      \includegraphics[width=2.8in]{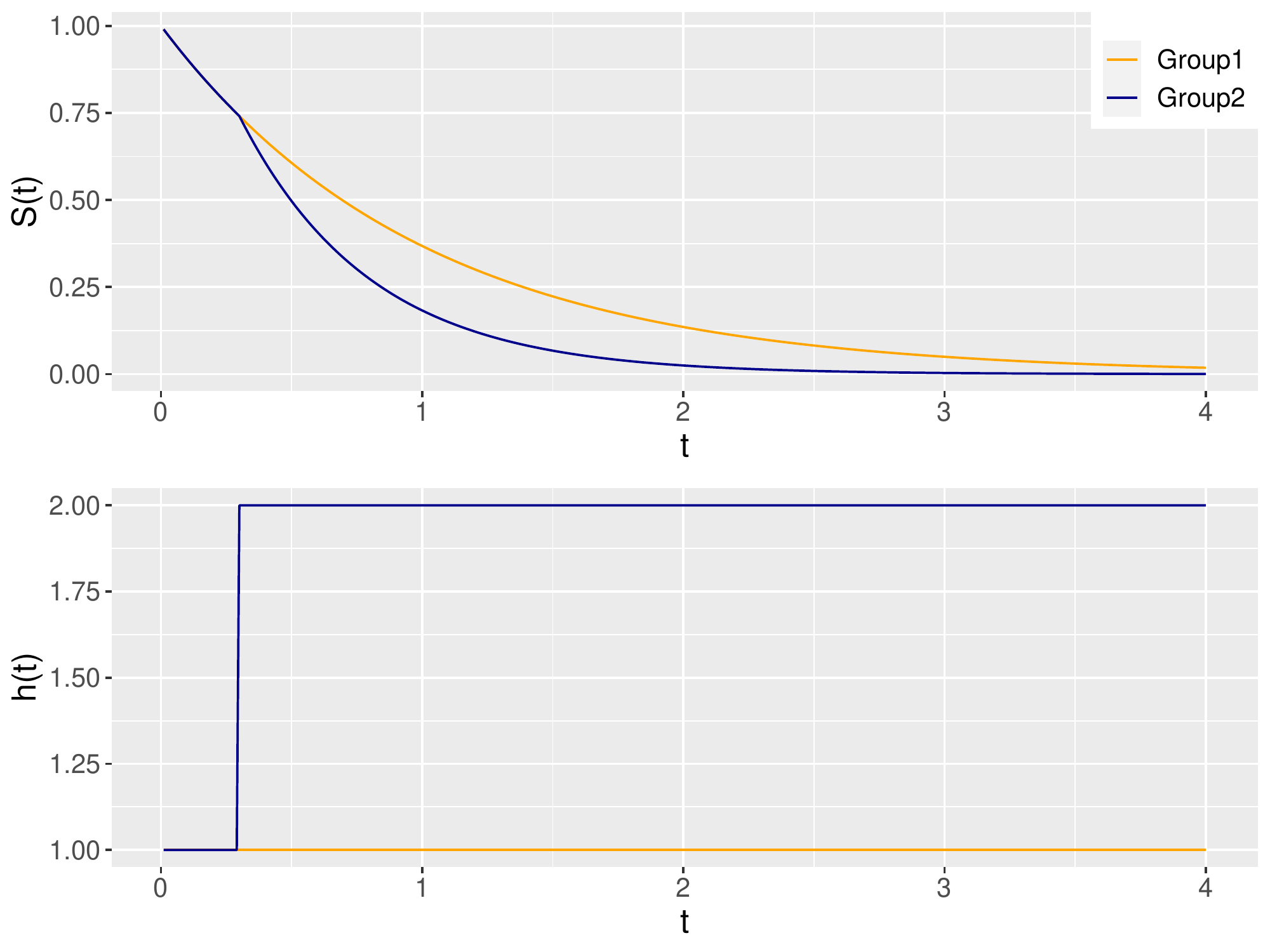}} \\
      \hline
      NPH3 &  {$\!\begin{aligned} 
                F_1(t) &= \textit{Exp}(1/28)\\
               F_2(t) &= \begin{cases}
                            \textit{Exp}(1/15) & \text{$t \leq 12$}\\
                            \textit{Exp}(1/28) & \text{$t > 12$}
                        \end{cases}   
                 \end{aligned}$} & \raisebox{-.5\height}{
      \includegraphics[width=2.8in]{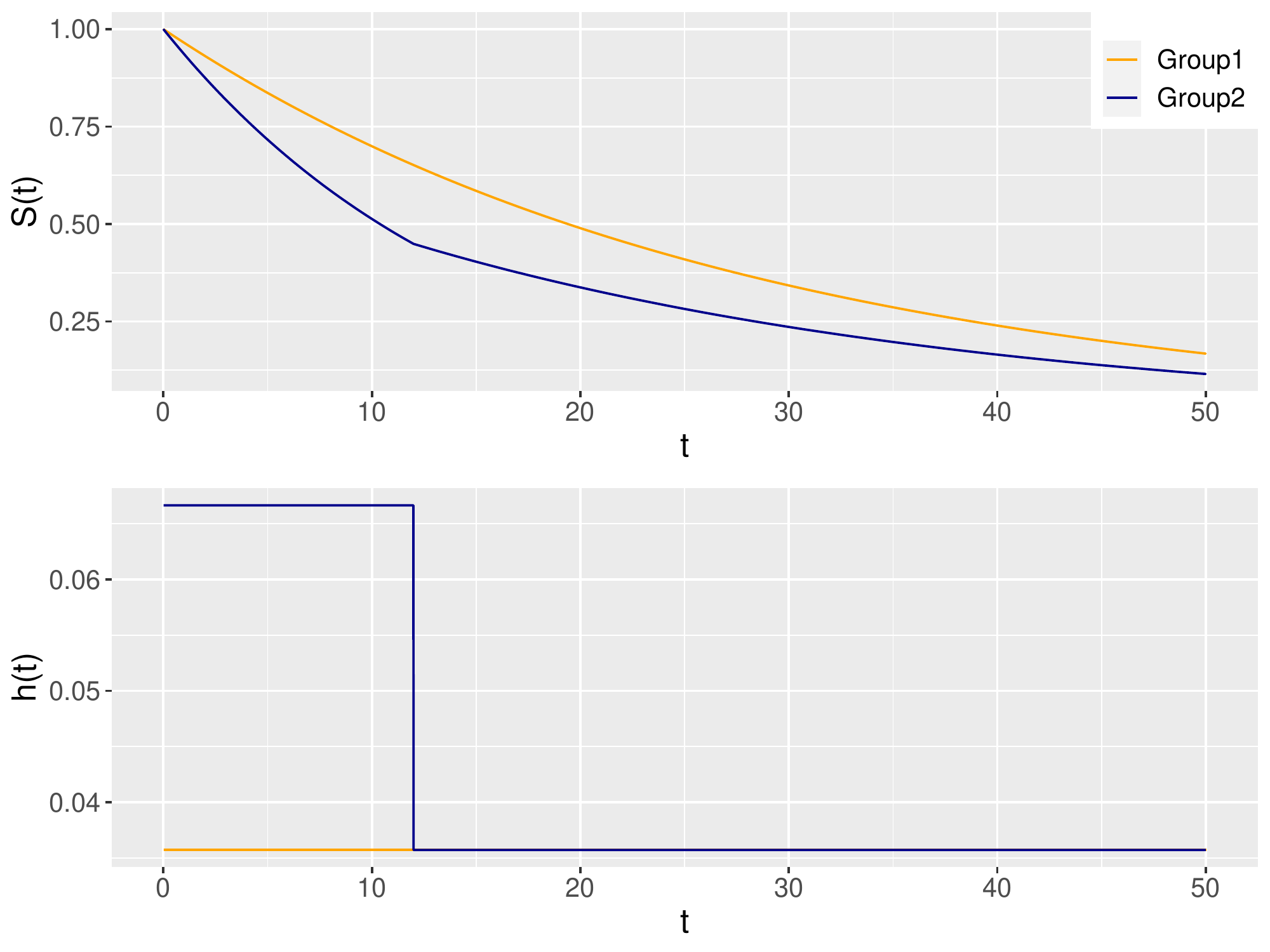}} \\
      \hline
      NPH4 &  {$\!\begin{aligned} 
               F_1(t) &= \textit{Lognormal}(1.2,1.7)\\  
               F_2(t) &= \textit{Lognormal}(2.4,1.6) \end{aligned}$} & \raisebox{-.5\height}{
      \includegraphics[width=2.8in]{Plots/LogN_new.pdf}} \\
      \hline
    \end{tabular}
    \caption{Scenarios with non-proportional hazard functions.}
    \label{tab:NProp}
\end{table}

\begin{table}[H]
    \centering
    \begin{tabular}{|c|c c|}
    \hline
    Scenario & CDF & Visualization of the survival and hazard curves \\
    \hline
         C1 \cite{gorfineKsampleOmnibusNonproportional2019} &  {$\!\begin{aligned} 
               F_1(t) &= \textit{Weibull}(0.849, 10)\\
               F_2(t) &= \begin{cases}
                            \textit{Weibull}(0.849,10) & \text{$t \leq 3$}\\
                            \textit{Unif}(3, 50.625) & \text{$3 < t \leq 33$}\\
                            \textit{Weibull}(0.849,10) & \text{$t > 33$}
                        \end{cases}   
                 \end{aligned}$} & \raisebox{-.5\height}{
      \includegraphics[width=2.8in]{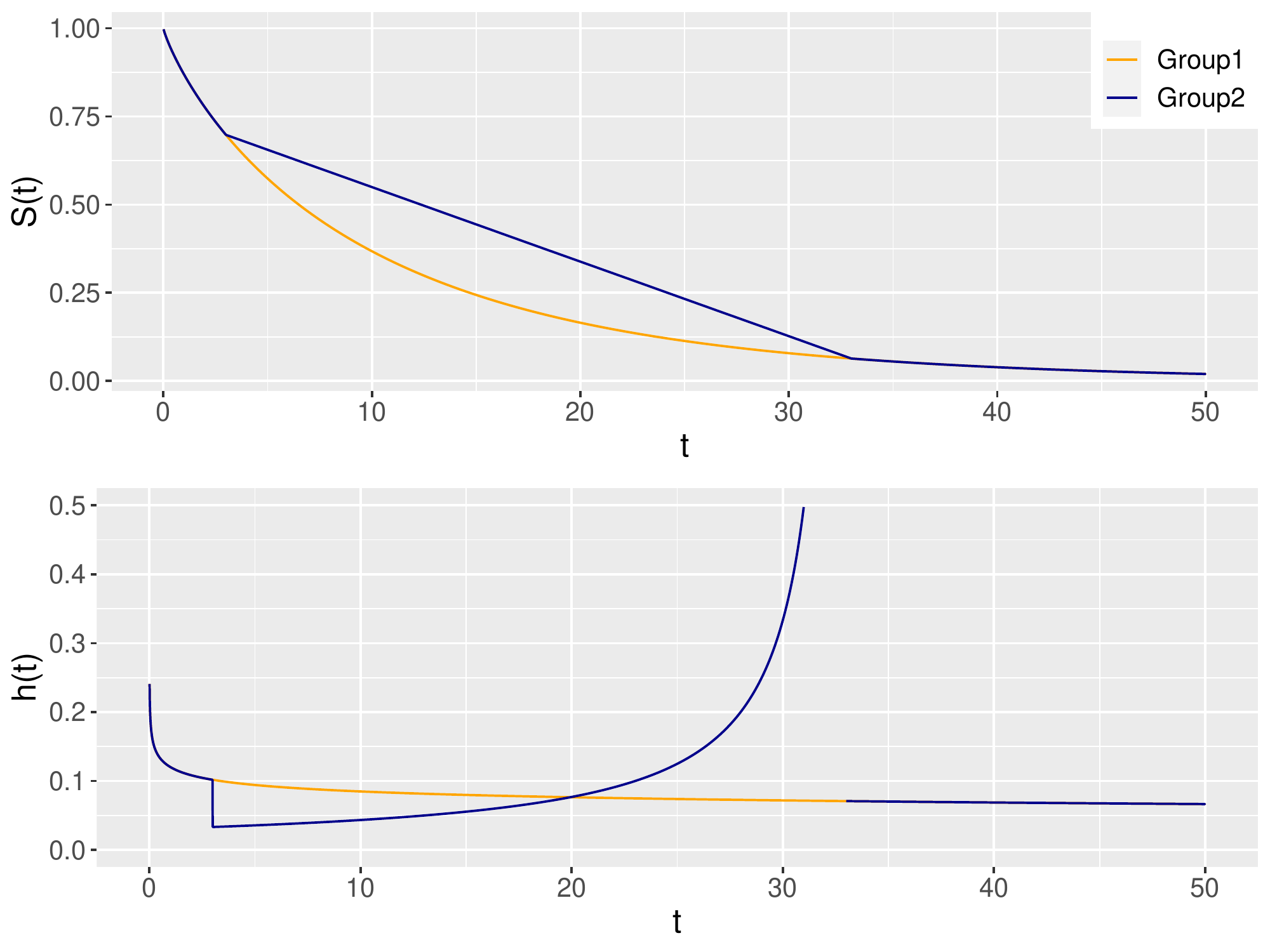}} \\ \\
      \hline
      C2 \cite{liStatisticalInferenceMethods2015} &  {$\!\begin{aligned} 
               F_1(t) &= \textit{Weibull}(2.5,30)\\
               F_2(t) &= \begin{cases}
                            \textit{Exp}(0.125) & \text{$t \leq 1$}\\
                            \textit{Exp}(0.01) & \text{$t > 1$}
                        \end{cases}    
                 \end{aligned}$} & \raisebox{-.5\height}{
      \includegraphics[width=2.8in]{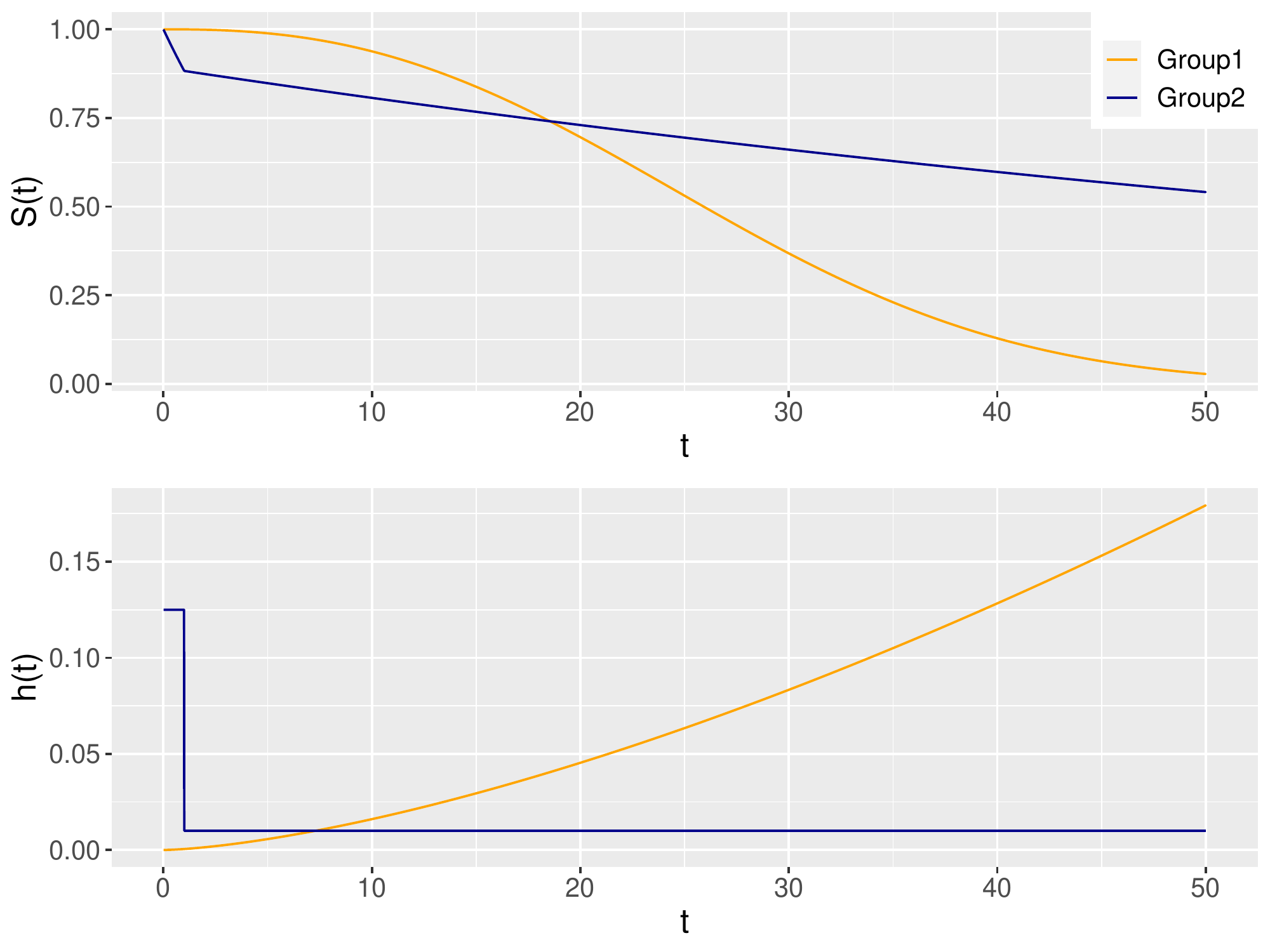}} \\
      \hline
      C3 \cite{liStatisticalInferenceMethods2015} &  {$\!\begin{aligned} 
               F_1(t) &= \textit{Exp}(1/12)\\
               F_2(t) &= \begin{cases}
                            \textit{Exp}(0.25) & \text{$t \leq 2$}\\
                            \textit{Exp}(1/35) & \text{$t > 2$}
                        \end{cases}    
                 \end{aligned}$} & \raisebox{-.5\height}{
      \includegraphics[width=2.8in]{Plots/Li3.pdf}} \\
      \hline
      C4 \cite{liStatisticalInferenceMethods2015} &  {$\!\begin{aligned} 
                F_1(t) &= \textit{Weibull}(1.5,5) \\
               F_2(t) &= \begin{cases}
                            \textit{Exp}(0.5) & \text{$t \leq 1.5$}\\
                            \textit{Exp}(0.1) & \text{$t > 1.5$}
                        \end{cases}  
               \end{aligned}$} & \raisebox{-.5\height}{
      \includegraphics[width=2.8in]{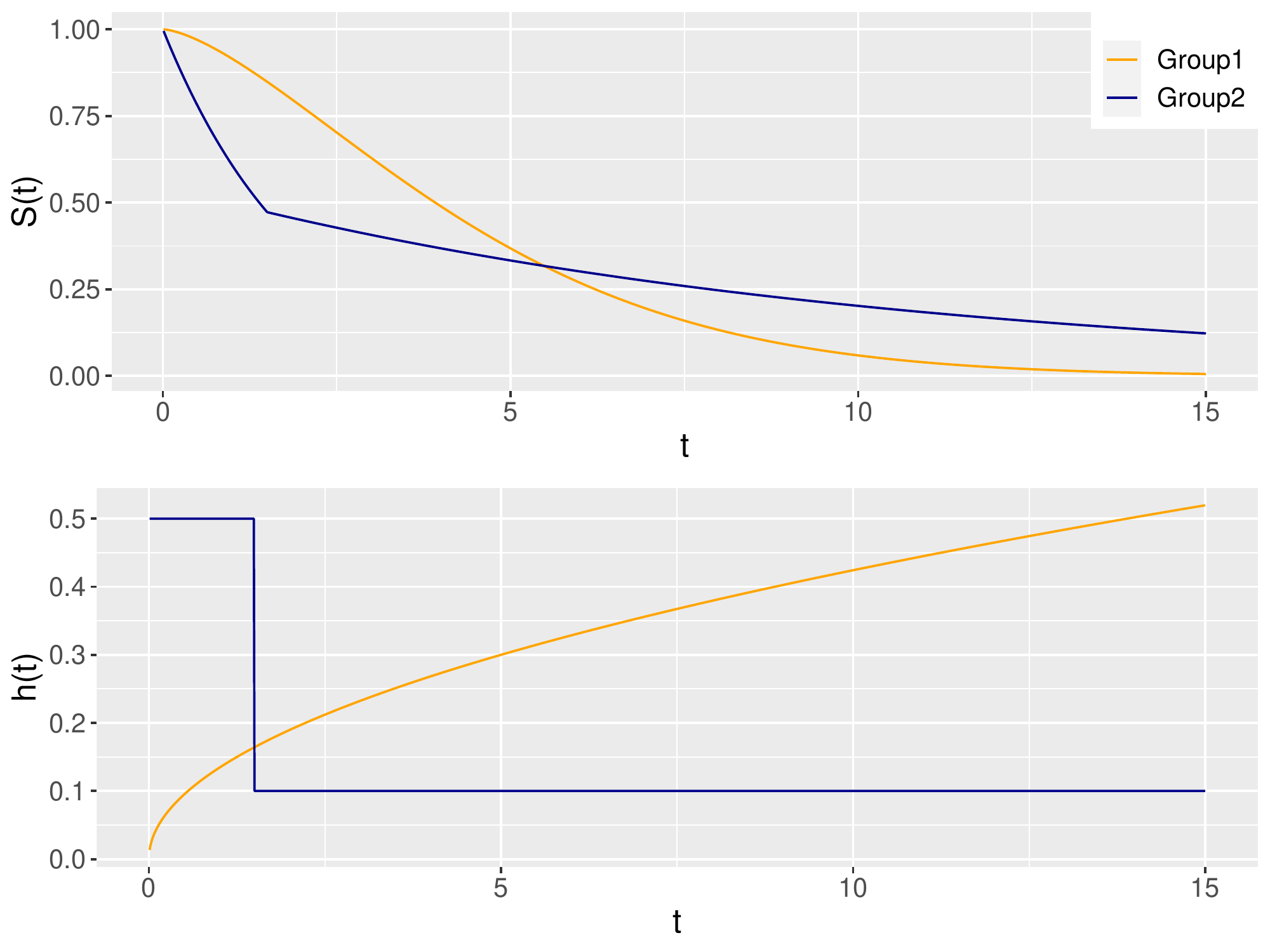}} \\
      \hline
    \end{tabular}
    \caption{Scenarios with crossing hazard functions.}
    \label{tab:Cross1}
\end{table}

\begin{table}[H]
    \centering
    \begin{tabular}{|c|c c|}
    \hline
    Scenario & CDF & Visualization of the survival and hazard curves \\
    \hline
         C5 &  {$\!\begin{aligned} 
               F_1(t) &= \textit{Gompertz}(0.2, 0.04) \\
               F_2(t) &= \textit{Gompertz}(0.07, 0.06)  \end{aligned}$} & \raisebox{-.5\height}{
      \includegraphics[width=2.8in]{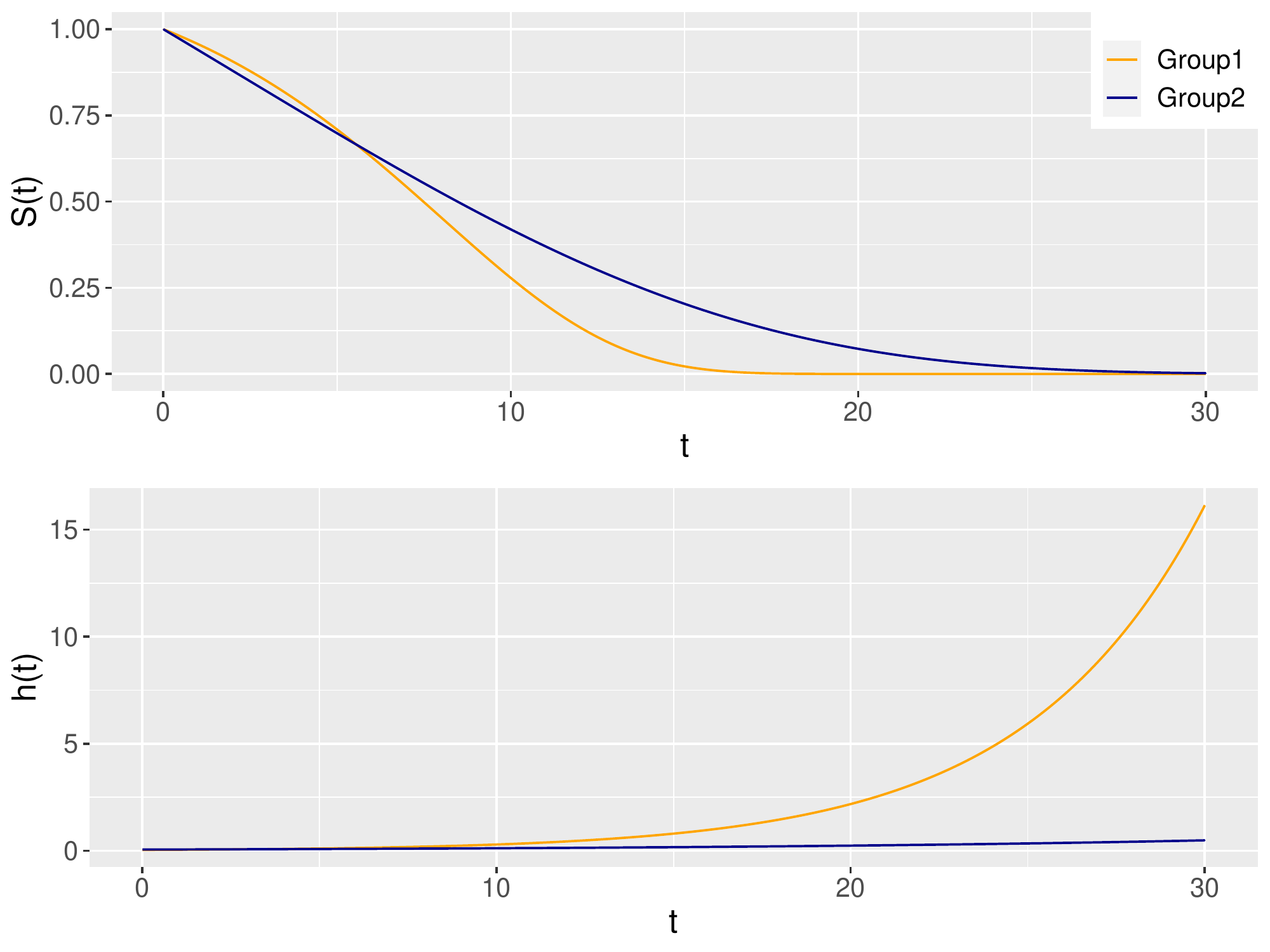}} \\
      \hline
      C6 \cite{gorfineKsampleOmnibusNonproportional2019} &  {$\!\begin{aligned} 
               F_1(t) &= \textit{Exp}(1) \\
               F_2(t) &= \begin{cases}
                            \textit{Exp}(2) & \text{$t \leq 0.25$}\\
                            \textit{Exp}(0.6) & \text{$t > 0.25$}
                        \end{cases}  
                 \end{aligned}$} & \raisebox{-.5\height}{
      \includegraphics[width=2.8in]{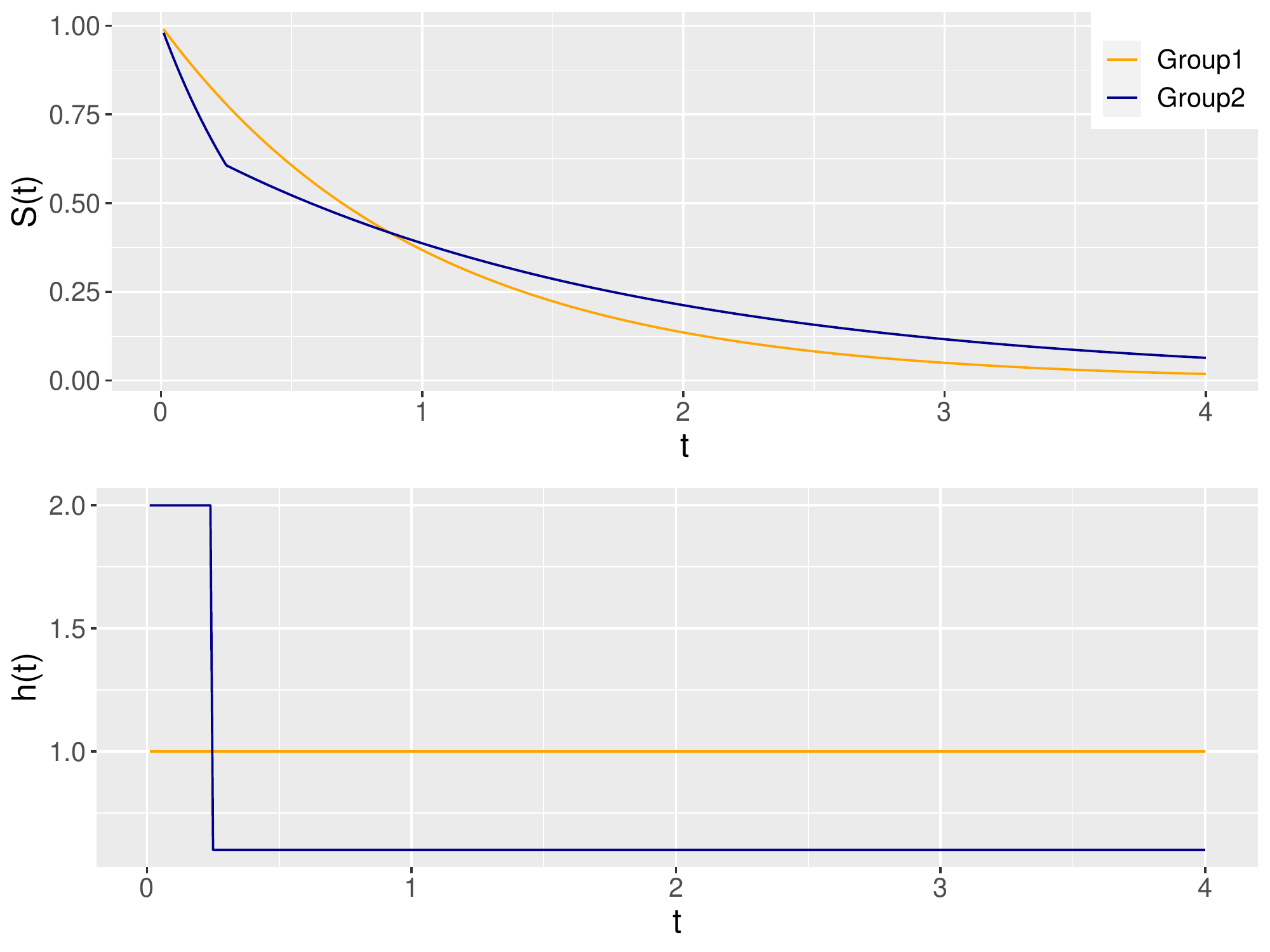}} \\
      \hline
      C7 &  {$\!\begin{aligned} 
               F_1(t) &= \textit{Exp}(0.1) \\   
               F_2(t) &= \textit{Weibull}(3,10)  \end{aligned}$} & \raisebox{-.5\height}{
      \includegraphics[width=2.8in]{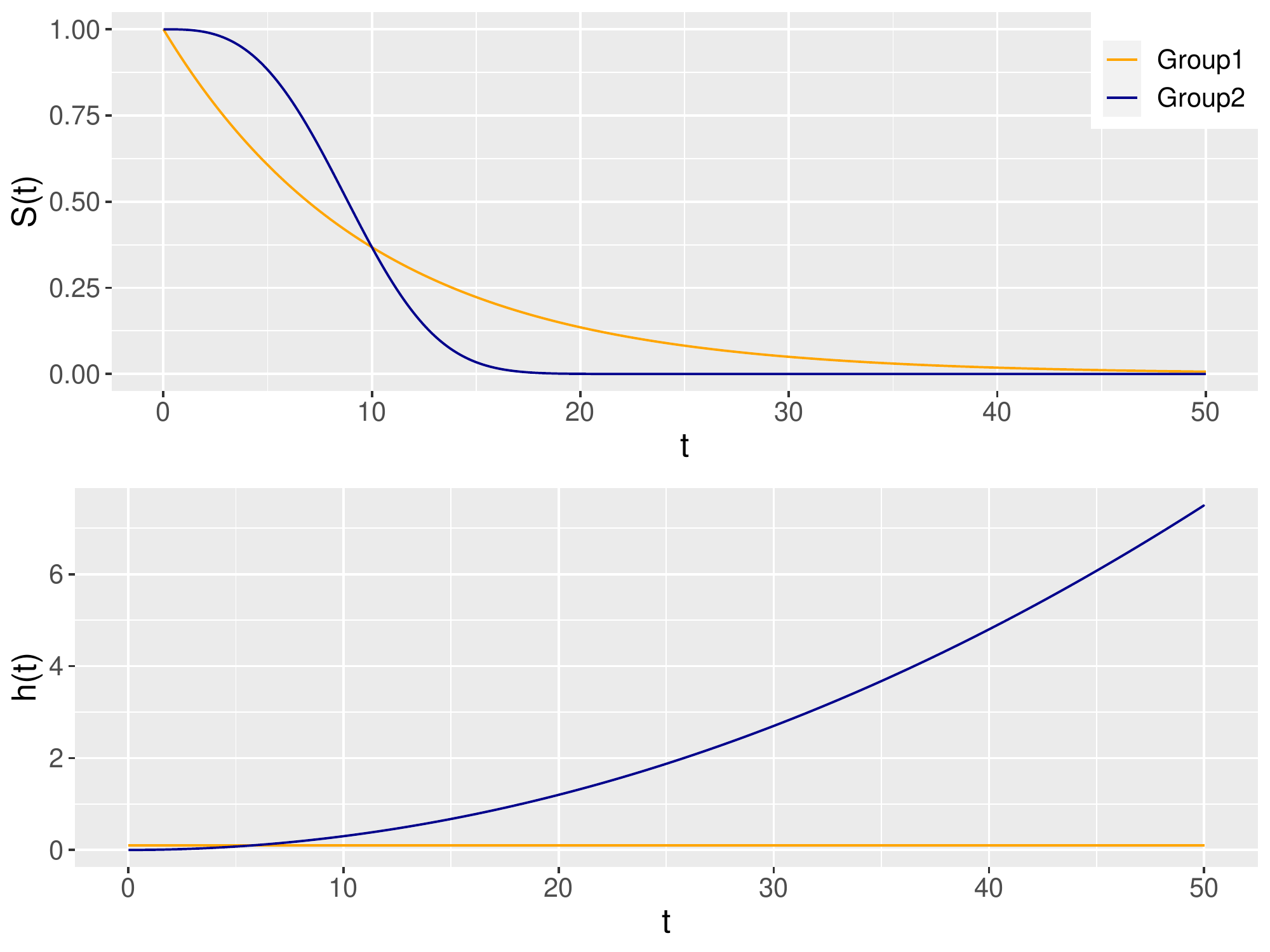}} \\
      \hline
      C8 &  {$\!\begin{aligned} 
               F_1(t) &= \textit{Weibull}(1.5,30) \\  
               F_2(t) &= \textit{Weibull}(3,25) \end{aligned}$} & \raisebox{-.5\height}{
      \includegraphics[width=2.8in]{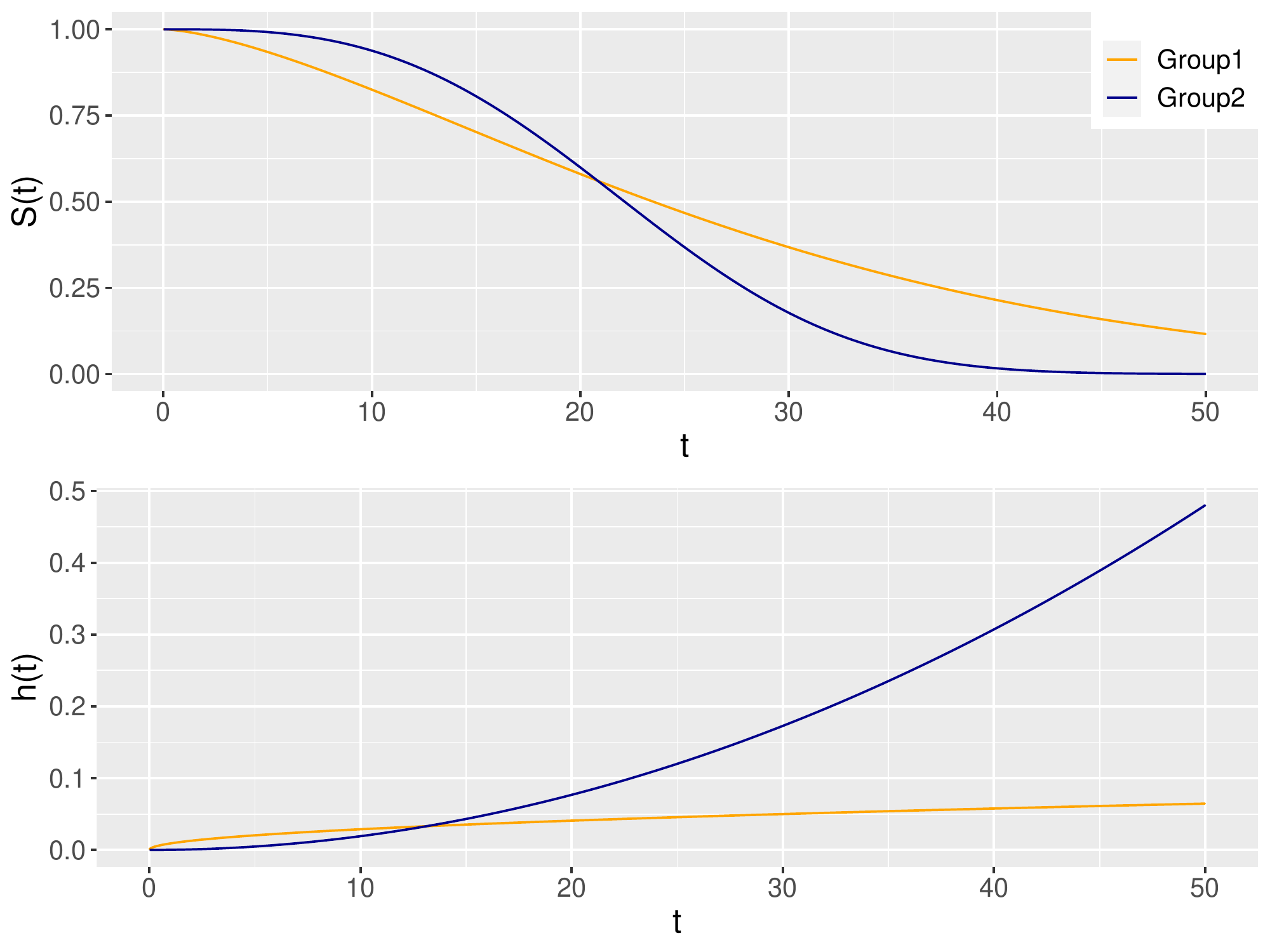}} \\
      \hline
    \end{tabular}
    \caption{Scenarios with crossing hazard functions.}
    \label{tab:Cross2}
\end{table}

\section*{Simulation results}

\begin{figure}[H]
    \centering
    \includegraphics[width=\textwidth]{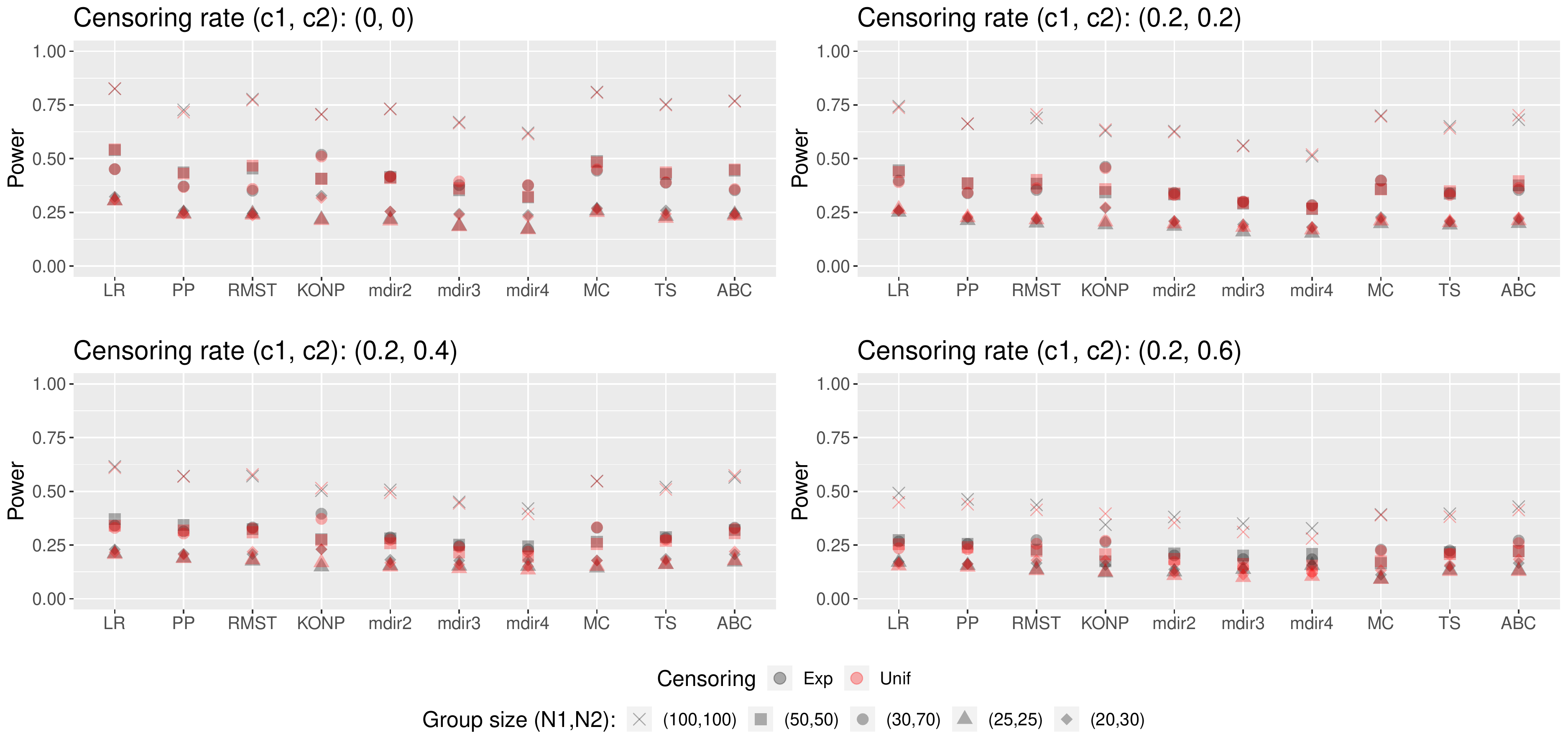}
    \caption{Simulation results Prop1}
    \label{fig:resProp1}
\end{figure}

\begin{figure}[H]
    \centering
    \includegraphics[width=\textwidth]{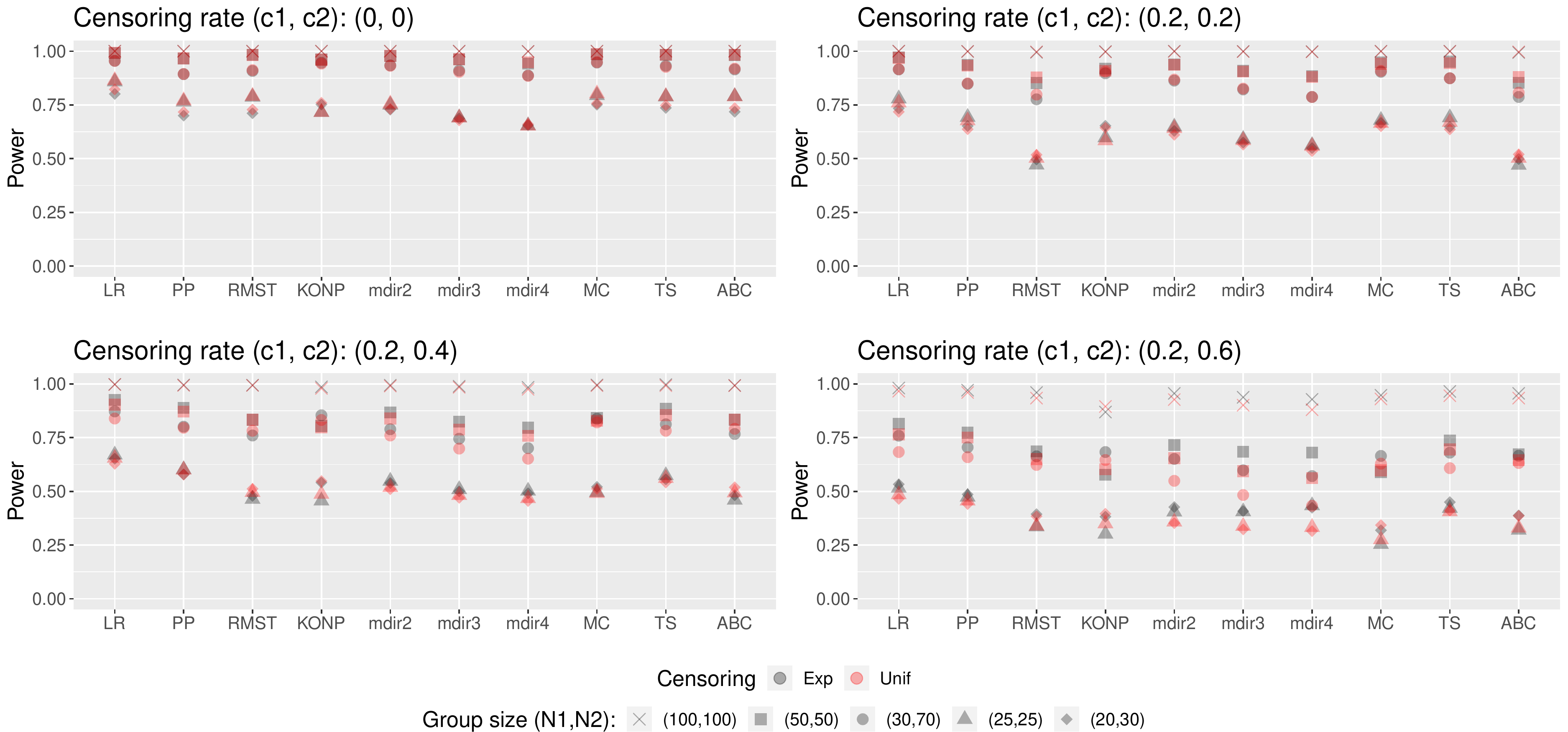}
    \caption{Simulation results Prop2}
    \label{fig:resProp2}
\end{figure}

\begin{figure}[H]
    \centering
    \includegraphics[width=\textwidth]{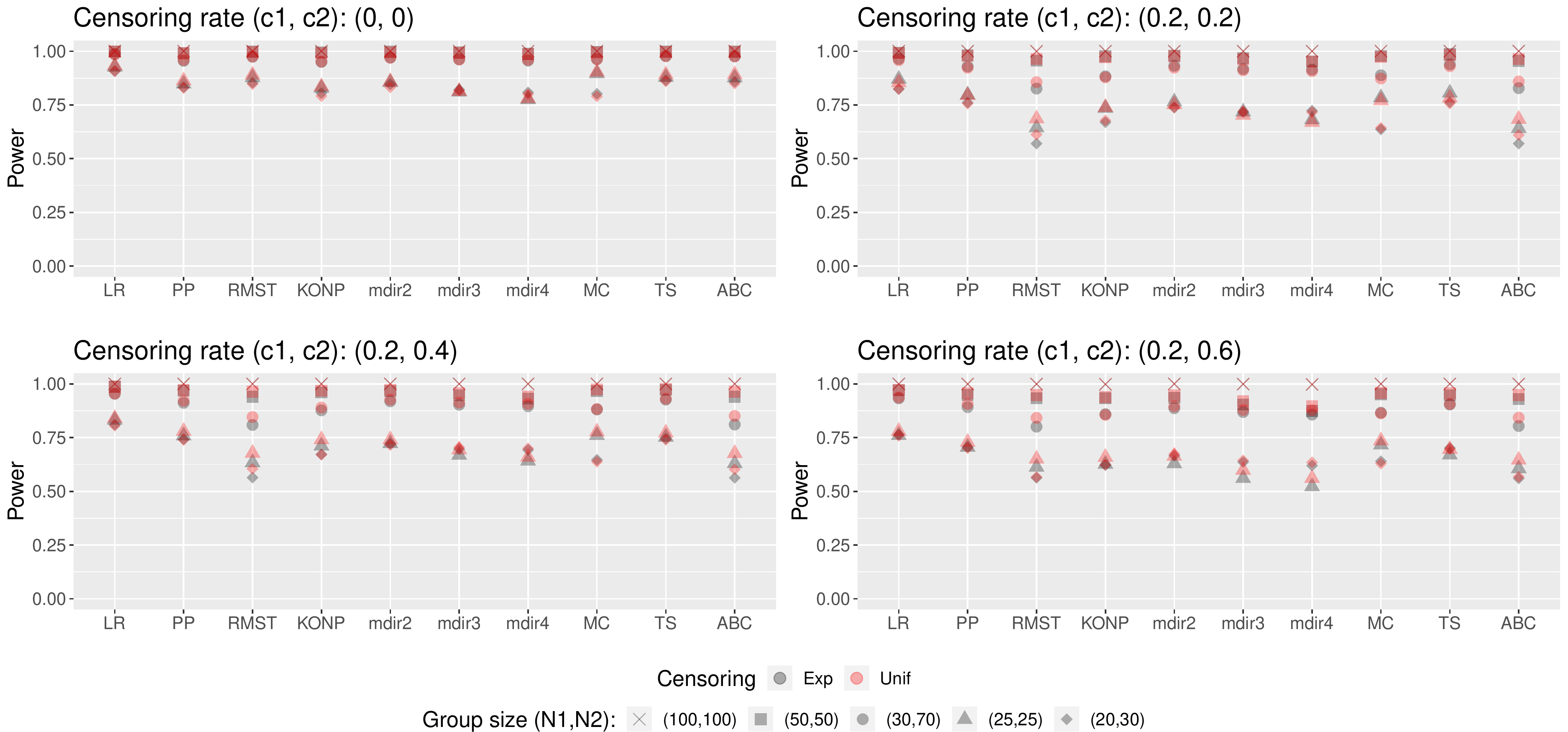}
    \caption{Simulation results Prop3}
    \label{fig:resProp3}
\end{figure}

\begin{figure}[H]
    \centering
    \includegraphics[width=\textwidth]{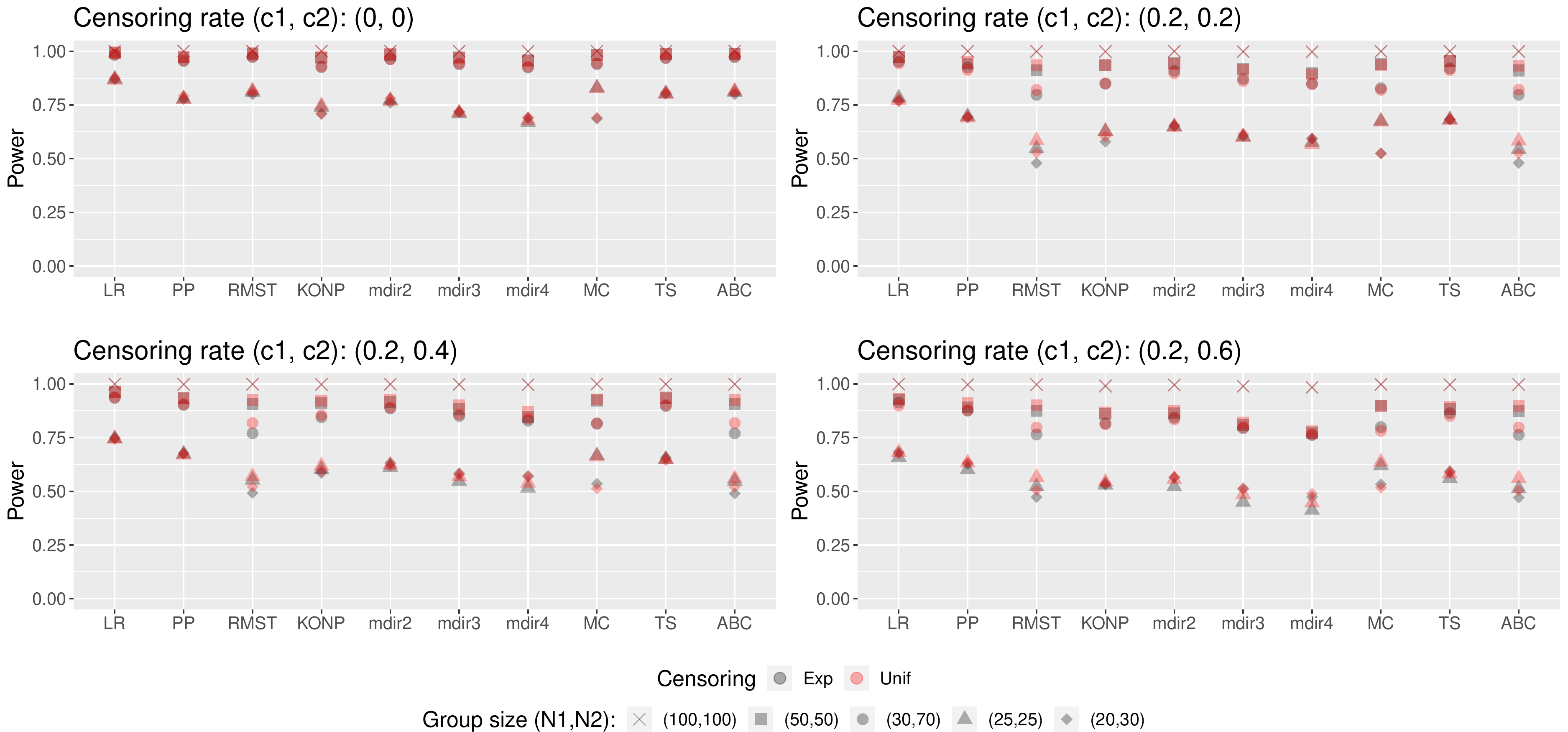}
    \caption{Simulation results Prop4}
    \label{fig:resProp4}
\end{figure}

\begin{figure}[H]
    \centering
    \includegraphics[width=\textwidth]{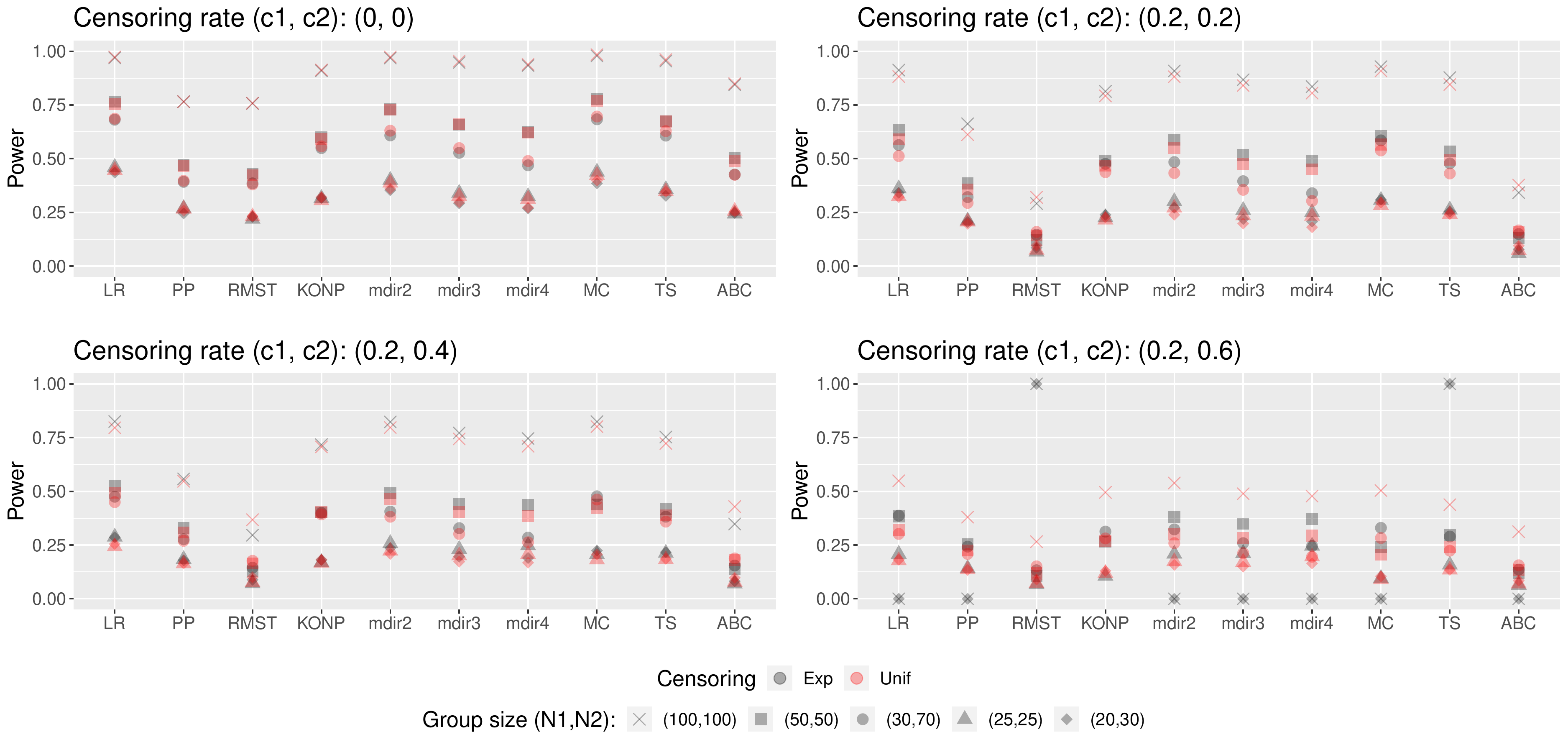}
    \caption{Simulation results NProp1}
    \label{fig:resNProp1}
\end{figure}

\begin{figure}[H]
    \centering
    \includegraphics[width=\textwidth]{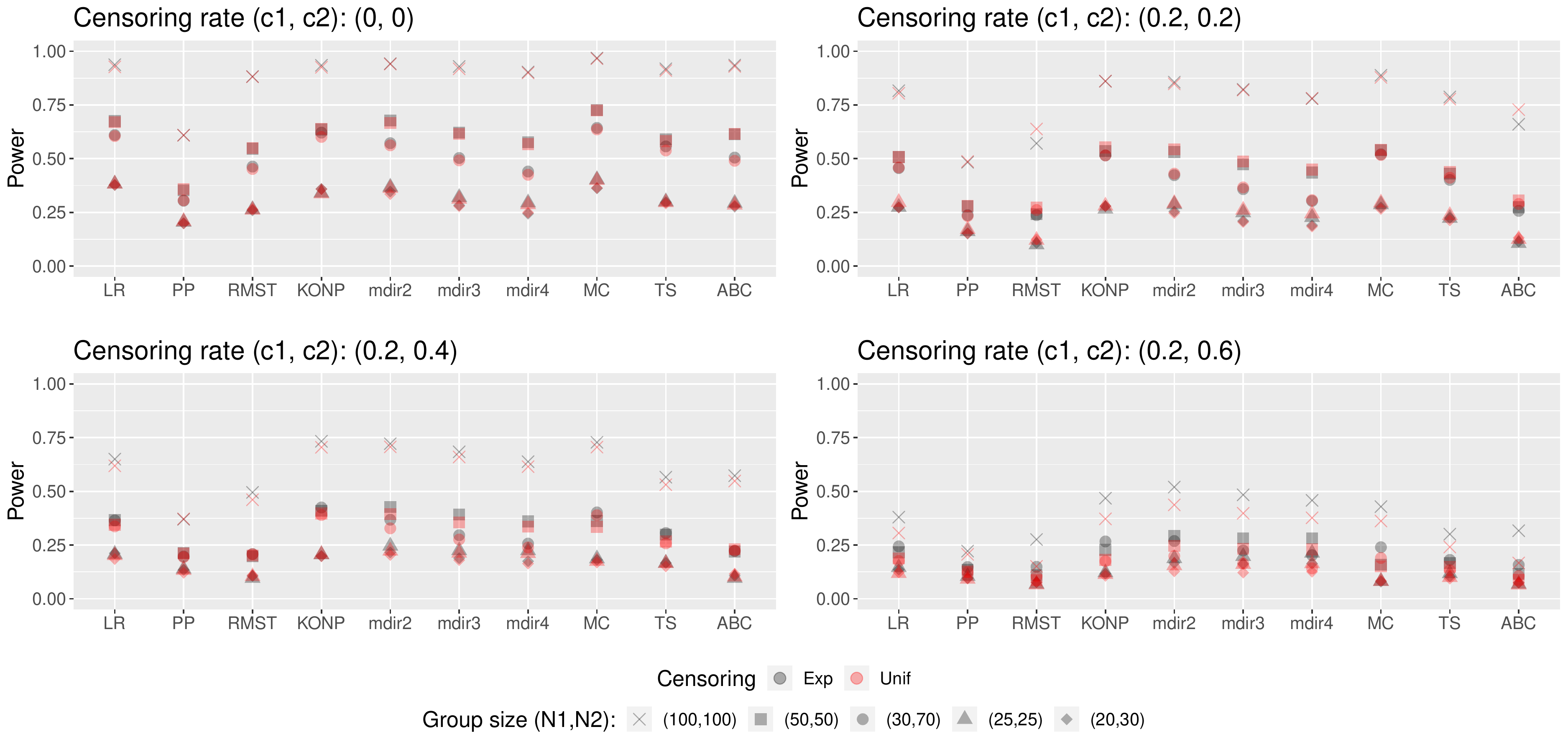}
    \caption{Simulation results NProp2}
    \label{fig:resNProp2}
\end{figure}

\begin{figure}[H]
    \centering
    \includegraphics[width=\textwidth]{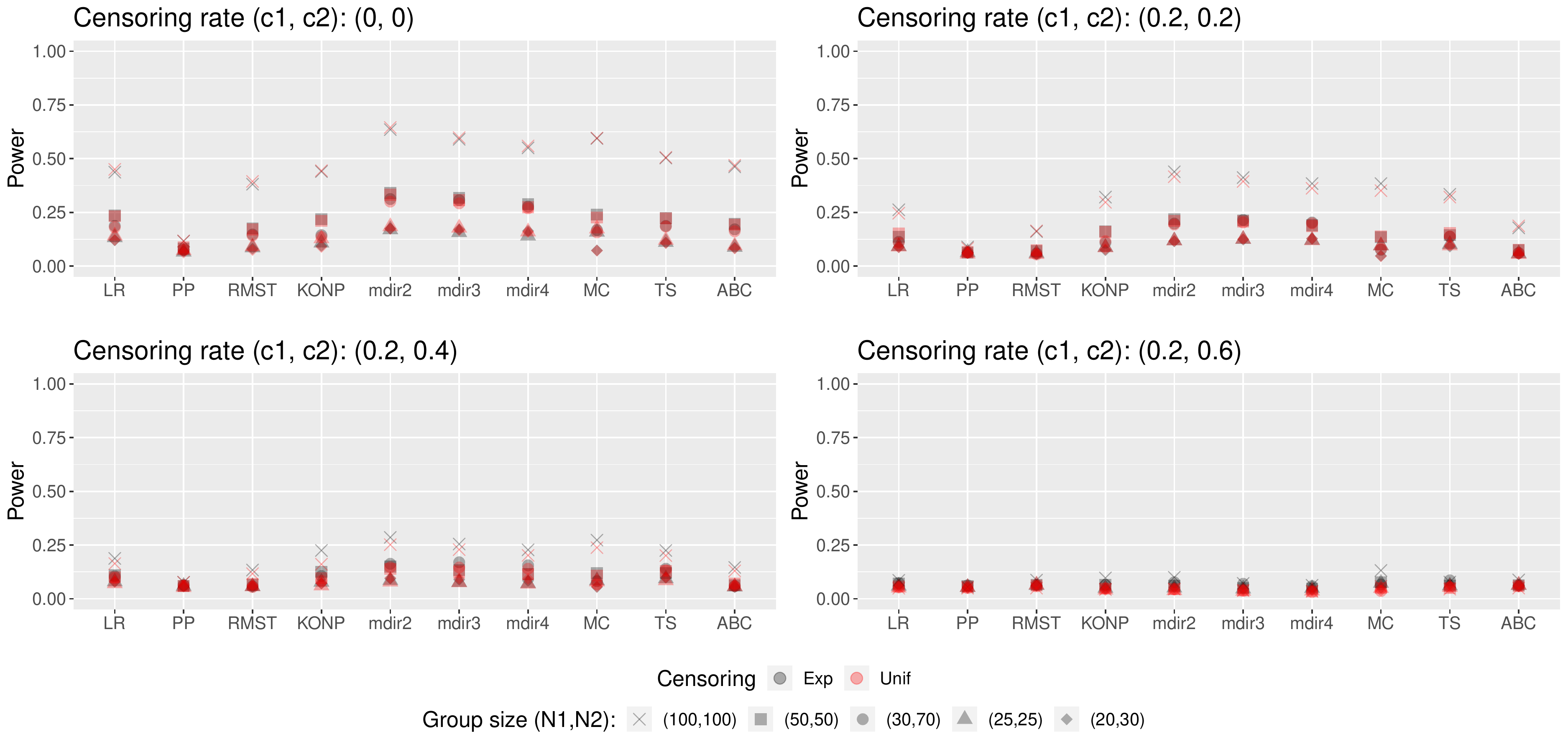}
    \caption{Simulation results NProp3}
    \label{fig:resNProp3}
\end{figure}

\begin{figure}[H]
    \centering
    \includegraphics[width=\textwidth]{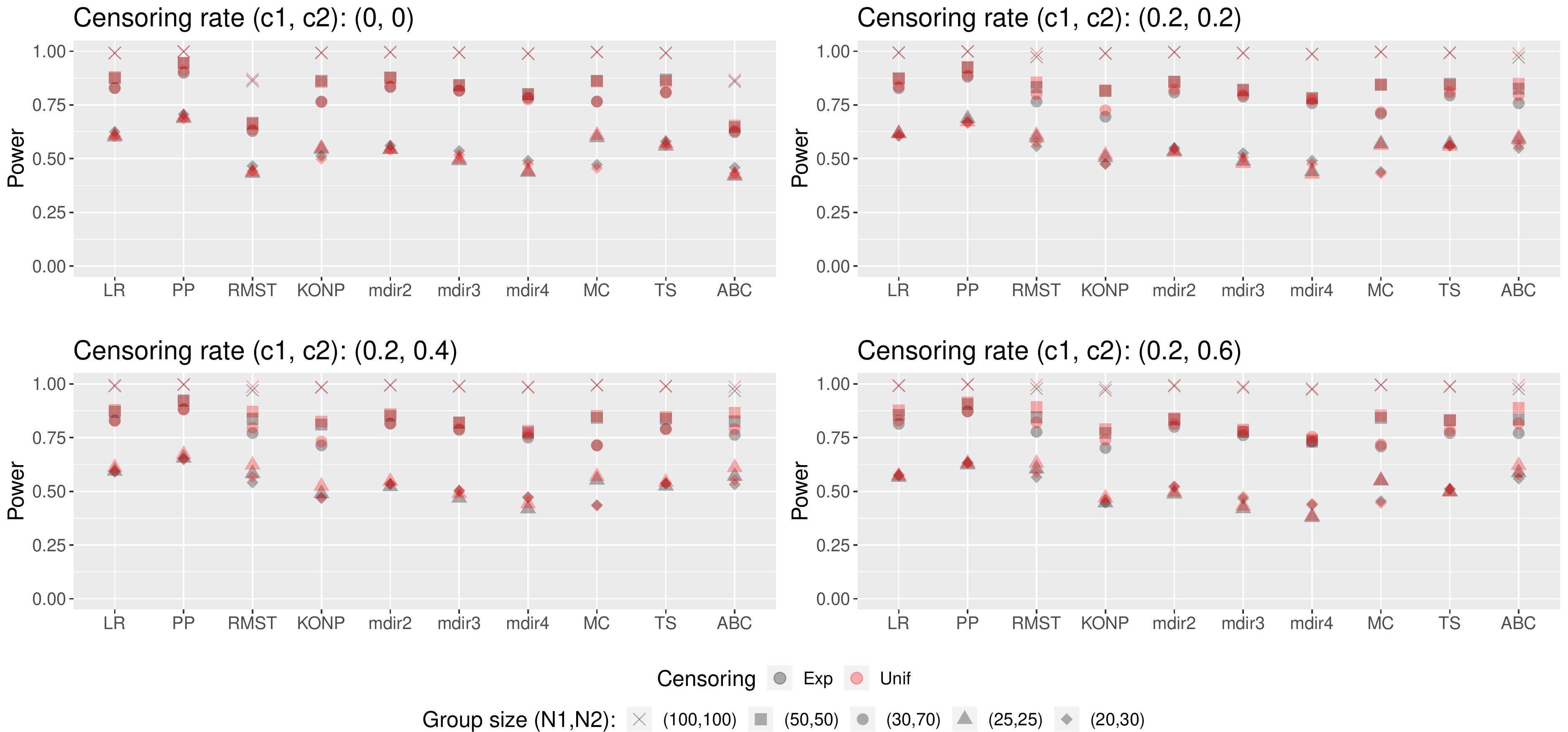}
    \caption{Simulation results NProp4}
    \label{fig:resNProp4}
\end{figure}

\begin{figure}[H]
    \centering
    \includegraphics[width=\textwidth]{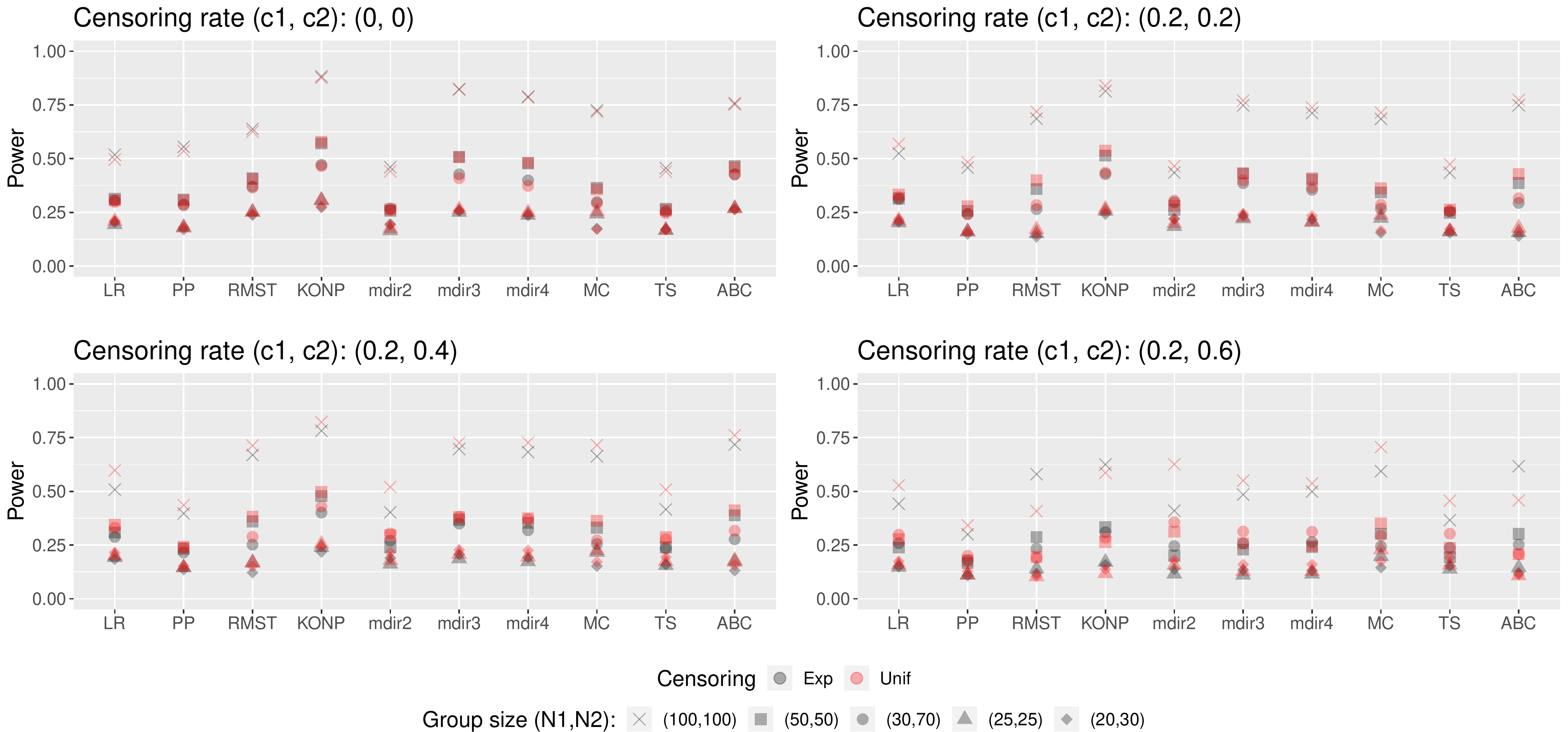}
    \caption{Simulation results Cross1}
    \label{fig:resCross1}
\end{figure}

\begin{figure}[H]
    \centering
    \includegraphics[width=\textwidth]{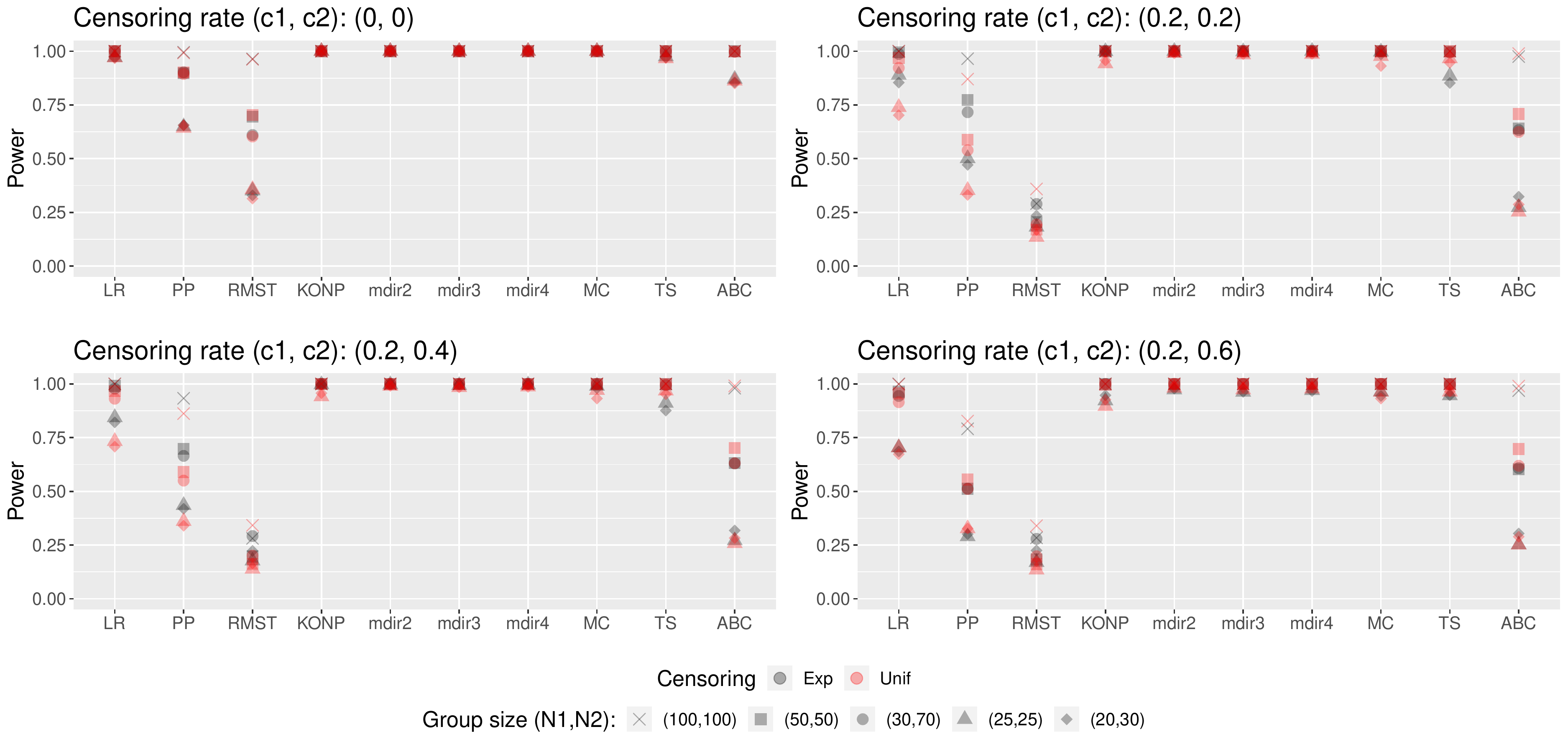}
    \caption{Simulation results Cross2}
    \label{fig:resCross2}
\end{figure}

\begin{figure}[H]
    \centering
    \includegraphics[width=\textwidth]{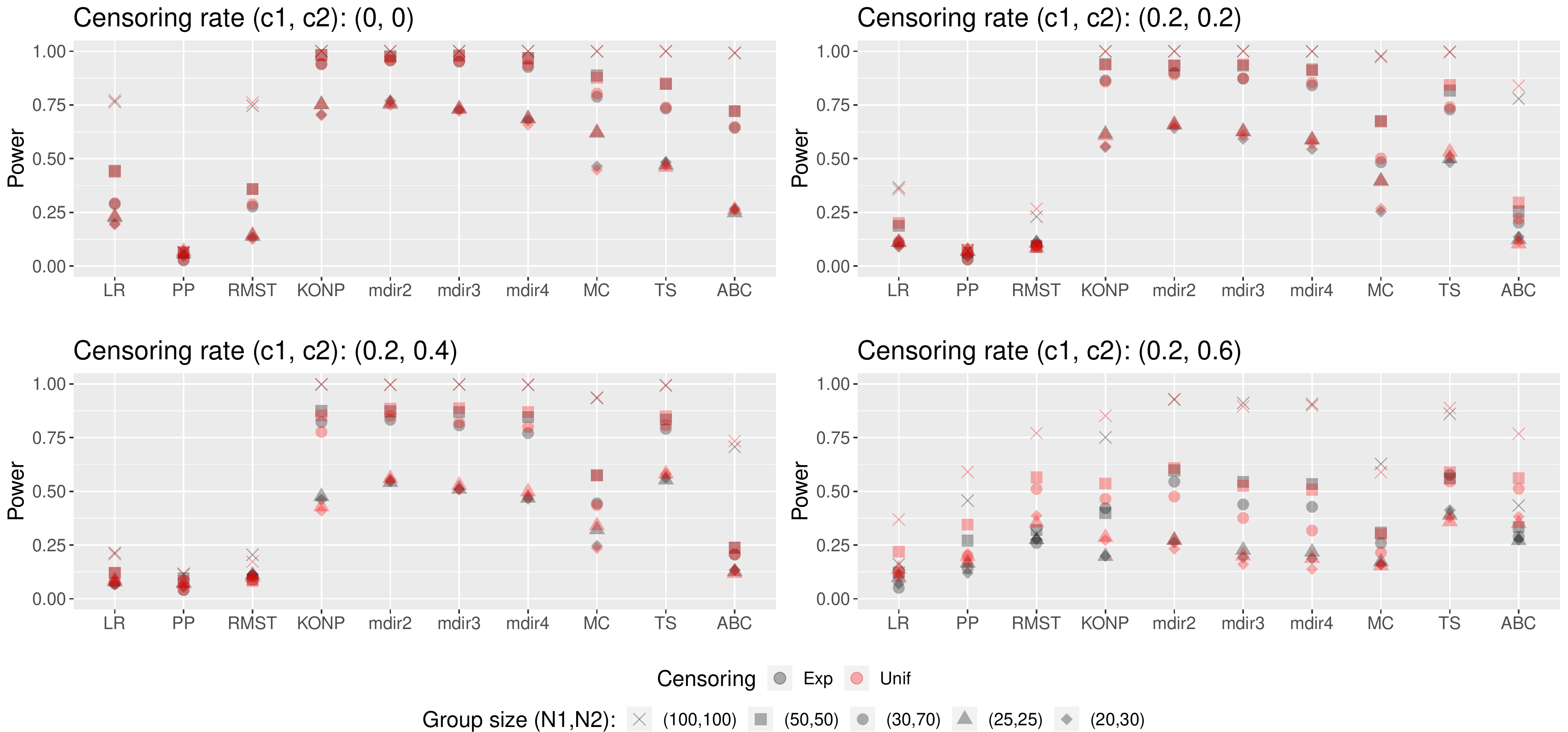}
    \caption{Simulation results Cross3}
    \label{fig:resCross3}
\end{figure}

\begin{figure}[H]
    \centering
    \includegraphics[width=\textwidth]{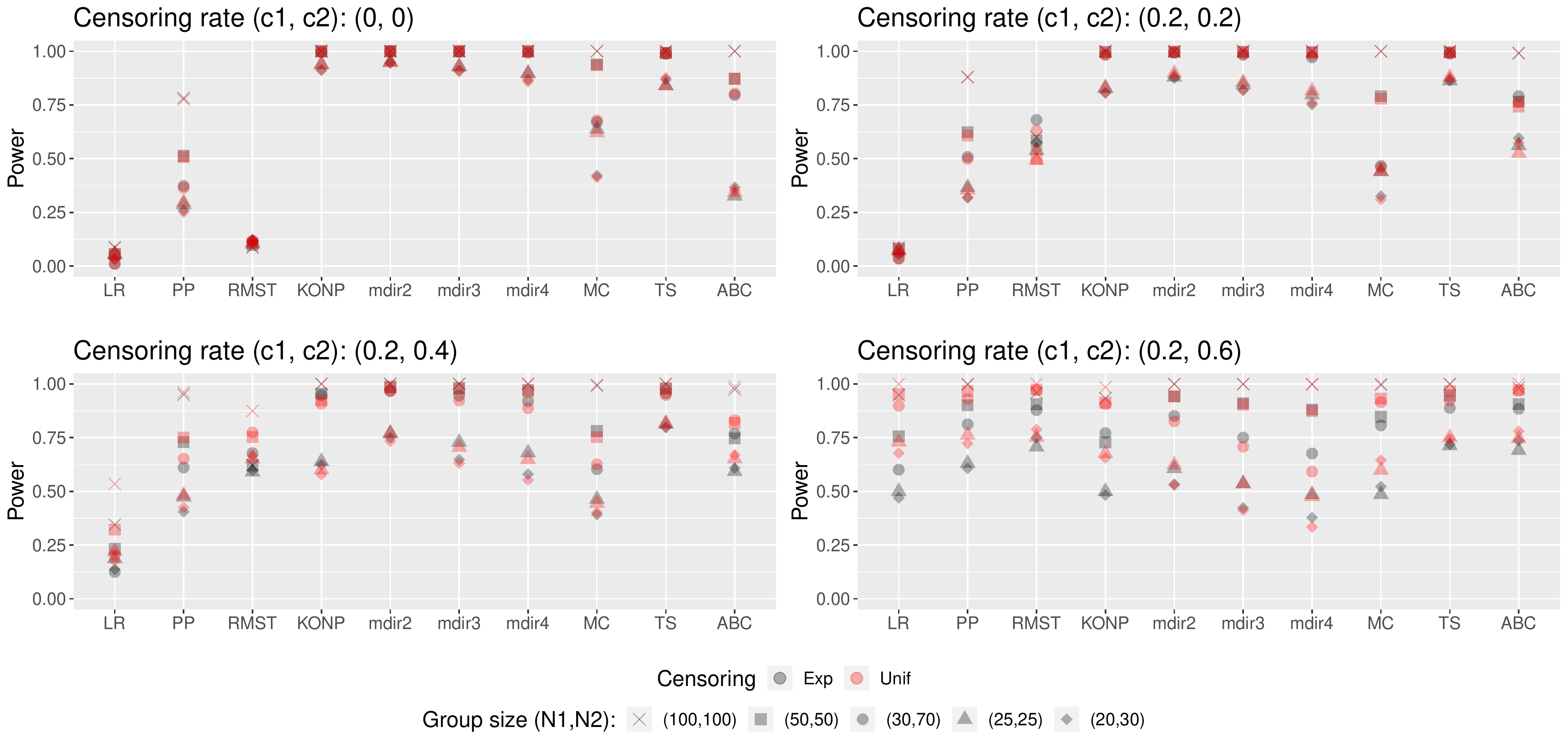}
    \caption{Simulation results Cross4}
    \label{fig:resCross4}
\end{figure}

\begin{figure}[H]
    \centering
    \includegraphics[width=\textwidth]{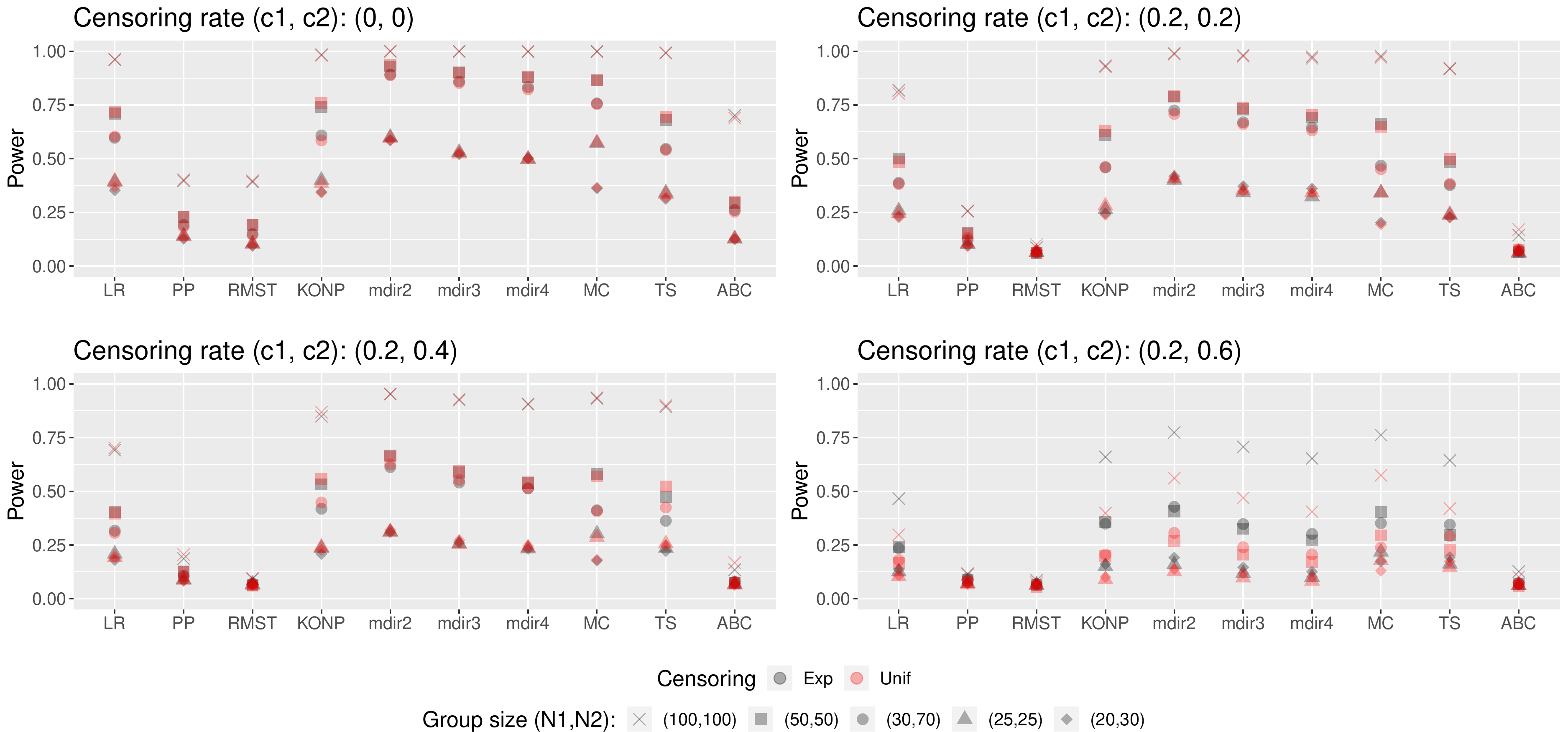}
    \caption{Simulation results Cross5}
    \label{fig:resCross5}
\end{figure}

\begin{figure}[H]
    \centering
    \includegraphics[width=\textwidth]{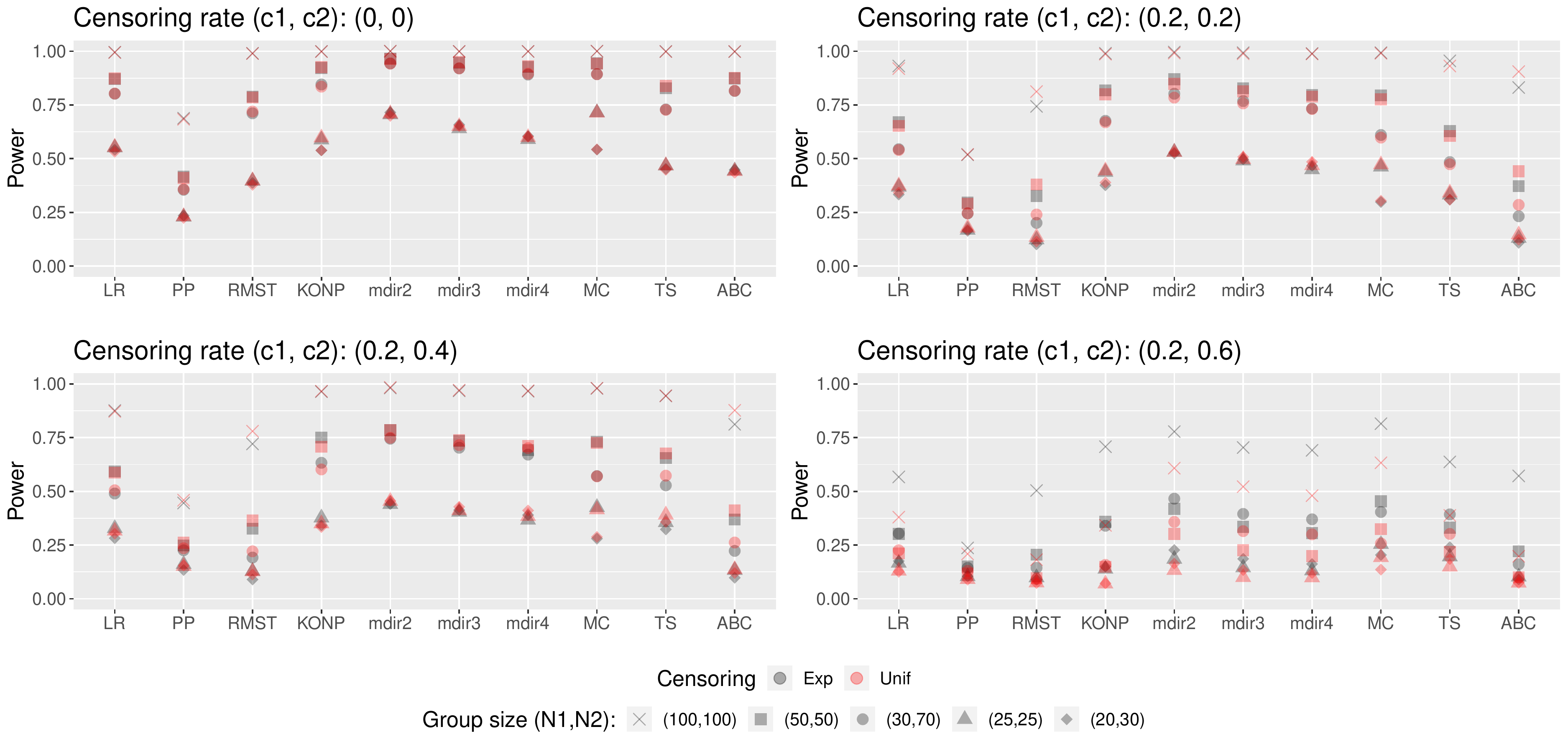}
    \caption{Simulation results Cross6}
    \label{fig:resCross6}
\end{figure}

\begin{figure}[H]
    \centering
    \includegraphics[width=\textwidth]{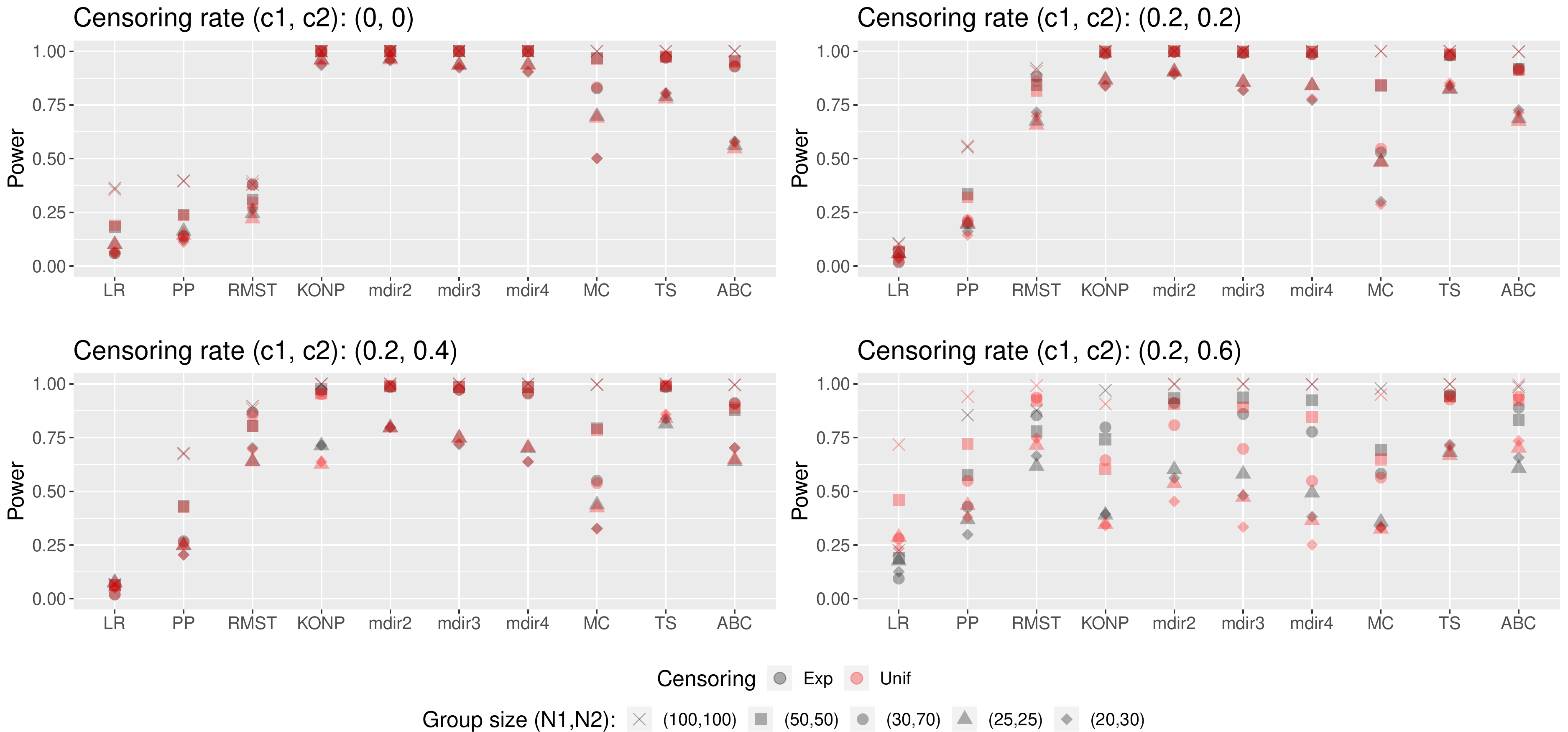}
    \caption{Simulation results Cross7}
    \label{fig:resCross7}
\end{figure}

\begin{figure}[H]
    \centering
    \includegraphics[width=\textwidth]{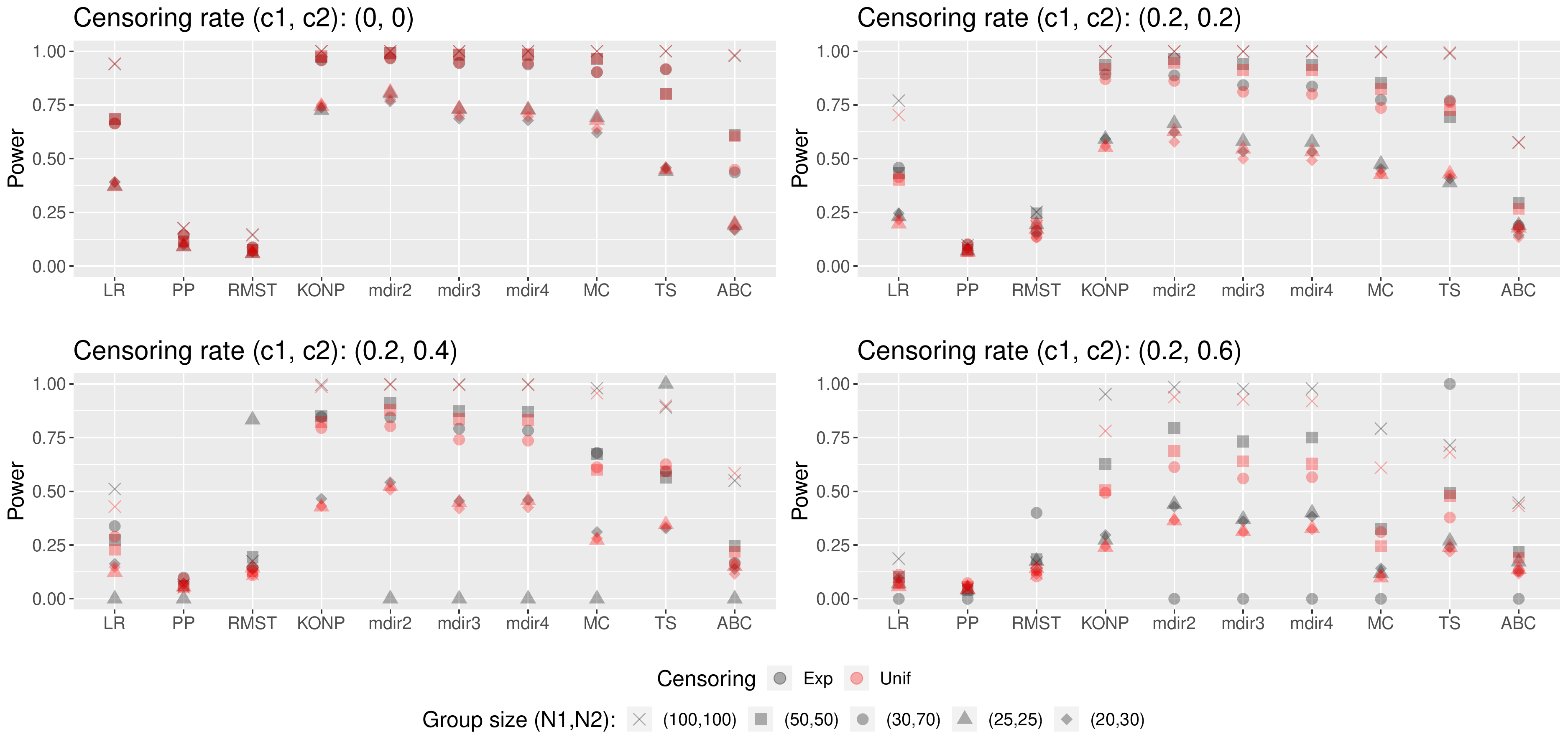}
    \caption{Simulation results Cross8}
    \label{fig:resCross8}
\end{figure}

\section*{Real Data Example}
In Table~\ref{tab:Jones} the published statistics as well as the reconstructed statistics are displayed to evaluate the reconstruction quality. Overall, it can be concluded that the reconstruction quality is sufficient. 

\begin{table}[H]
    \centering
    \begin{tabular}{|lcc|}
    \hline
    Statistic & Published data & Reconstructed data\\
    \hline
    Median progression free survival Dacarbazine & 8.10 & 8.02 \\
    Median progression free survival Trabectedin & 15.10 & 15.08\\
    Hazard ratio &0.72 (0.45, 1.17) & 0.71 (0.44, 1.15)\\
    \hline
    \end{tabular}
    \caption{Quality of data reconstruction.}
    \label{tab:Jones}
\end{table}